\documentclass[10pt,pra,twocolumn,superscriptaddress,showpacs,floatfix]{revtex4}
\usepackage{graphics}
\usepackage{graphicx}
\usepackage{bm}
\usepackage{amsmath}
\usepackage{amsfonts}
\usepackage{amssymb}
\usepackage{latexsym}
\usepackage{color}
\usepackage{bbold}
\usepackage{braket}

\usepackage{hyperref}

\begin{document}

\title{Quantising the electromagnetic field near two-sided semi-transparent mirrors}

\author{Nicholas Furtak-Wells}
\affiliation{The School of Physics and Astronomy, University of Leeds, Leeds LS2 9JT, United Kingdom}
\author{Lewis A. Clark}
\affiliation{The School of Physics and Astronomy, University of Leeds, Leeds LS2 9JT, United Kingdom}
\affiliation{Joint Quantum Centre Durham-Newcastle, School of Mathematics, Statistics and Physics, Newcastle University, Newcastle upon Tyne NE1 7RU, UK}
\author{Robert Purdy} 
\affiliation{The School of Physics and Astronomy, University of Leeds, Leeds LS2 9JT, United Kingdom}
\author{Almut Beige}
\affiliation{The School of Physics and Astronomy, University of Leeds, Leeds LS2 9JT, United Kingdom}
\date{\today}

\begin{abstract}
This paper models light scattering through flat surfaces with finite transmission, reflection and absorption rates, with wave packets approaching the mirror from both sides. While using the same notion of photons as in free space, our model also accounts for the presence of mirror images and the possible exchange of energy between the electromagnetic field and the mirror surface. To test our model, we derive the spontaneous decay rate and the level shift of an atom in front of a semi-transparent mirror as a function of its transmission and reflection rates. When considering limiting cases and using standard approximations, our approach reproduces well-known results but it also paves the way for the modelling of more complex scenarios.
\end{abstract}

\maketitle

\section{Introduction} \label{Intro}

The question, how to model the emission of light from atomic systems, is older than quantum physics itself. Planck's seminal paper on the spectrum of black body radiation \cite{Planck} is what eventually led to the discovery of quantum physics. Nowadays, we routinely use quantum optical master equations \cite{Agarwal2,He1,Stokes} to analyse the dynamics of atomic systems with spontaneous photon emission. For example, it is well known that the spontaneous decay rate of a two-level atom with ground state $|1 \rangle$ and excited state $|2 \rangle$ equals
\begin{eqnarray} \label{Gamma0}
\Gamma_{\rm free} &=& \frac{e^{2} \omega_0^{3} \| \mathbf{D}_{12} \|^{2}}{3 \pi \hbar \varepsilon c^{3}} 
\end{eqnarray} 
in a medium with permittivity $\varepsilon$. Here \textit{e} is the charge of a single electron and \textit{c} denotes the speed of light in the medium. Moreover, $\omega_{0}$ denotes the frequency and $\mathbf{D}_{12}$ is the dipole moment of the 1-2 transition. 

The spontaneous photon emission of atoms near perfect mirrors too has been extensively studied in the literature \cite{Morawitz,Stehle,Milonni,Kleppner,Sipe,Arnoldus1,Meschede,Drabe,Matloob,Dorner,Beige}. When considering this problem, boundary conditions of vanishing transversal electric and normal magnetic field amplitudes on the mirror surface need to be imposed. This is usually done by reducing the state space of the electromagnetic field to a subset of photon modes. Compared to free space, only half of the Hilbert space of the electromagnetic field is taken into account. As a result, the spontaneous decay rate $\Gamma_{\rm mirr}$ of an atom in front of a perfectly-reflecting mirror differs strongly from its free-space decay rate $\Gamma_{\rm free}$ in Eq.~\eqref{Gamma0}, when the distance $x$ of the atom from the mirror surface is of the same order of magnitude as the wavelength $\lambda_0$ of the emitted light. Although the effect of the mirror is relatively short-range, the sub- and super-radiance of atomic systems near perfect mirrors has already been verified experimentally \cite{Drexhage,Eschner,Delsing}.

\begin{figure}[b] 
	\centering
	\includegraphics[width=0.5\textwidth]{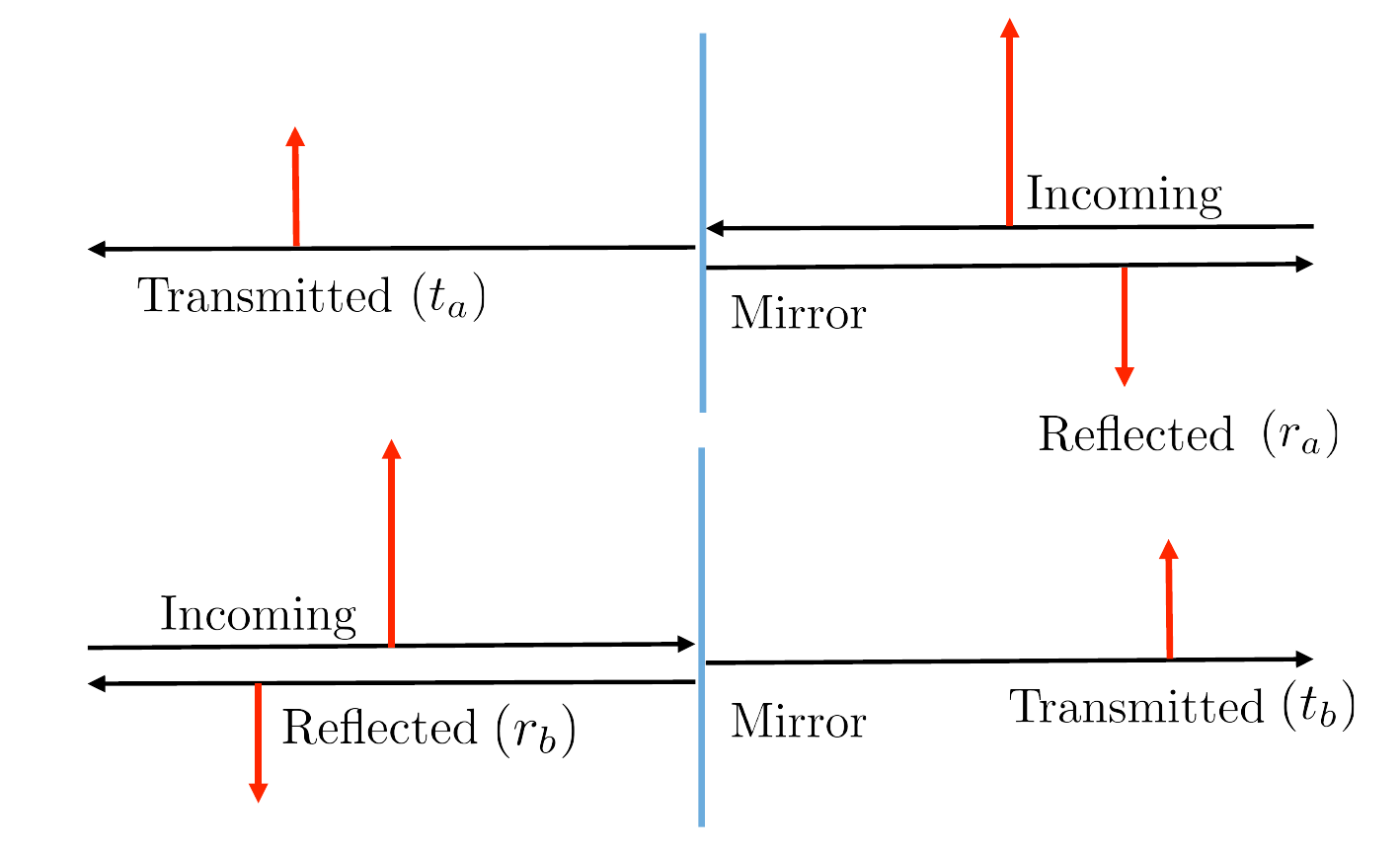} 
\caption{[Colour online] Schematic view of a semi-transparent mirror with light incident from both sides. Depending on the direction of the incoming light, we denote the (real) transmission and reflection rates of the mirror $t_{a}$, $t_{b}$, $r_a$ and $r_{b}$, respectively. In this paper, the possible absorption of light in the mirror surface is explicitly taken into account. However, for simplicity we assume that the medium on both sides of the mirror is the same.}
	\label{fig:intro}
\end{figure}

Quantising the electromagnetic field near semi-transparent mirrors is less straightforward. The foundation for this was laid by Sommerfeld in 1909, when he examined the propagation of surface waves above a flat lossy ground for applications in wireless communication \cite{Sommerfeld}. In 1971, Carniglia and Mandel \cite{Carniglia} considered a semi-transparent mirror with finite transmission and reflection rates and identified a set of elementary orthogonal light modes of travelling waves, so-called triplet modes. These are formed by incident, transmitted and reflected electromagnetic waves, as illustrated in Fig.~\ref{fig:intro}. Quantising these triplet modes, Carniglia and Mandel obtained an electromagnetic field Hamiltonian, which is the sum of independent harmonic oscillator Hamiltonians, and electromagnetic field observables, which are superpositions of free space observables (see e.g.~Refs.~\cite{Wu,Eberlein09,Eberlein12,Zakowicz,Hammer} for more recent related work).

In 1974, Agarwal used quantum electrodynamics to calculate the level shift and spontaneous decay rate of an atom near a dielectric medium \cite{Agar,Barton1}. Subsequently, he published a series of papers on quantum electrodynamics in the presence of dielectrics and conductors \cite{Agar1,Agar2,Agar3,Agar4,Agar5,Agar6}. In these papers, Agarwal uses linear response functions to indirectly deduce the properties of the electromagnetic field observables. His implicit approach to field quantisation helped to lay the foundations of a research area now known as macroscopic quantum electrodynamics \cite{Buhmann,Scheel1}. Other authors are more interested in the direct canonical quantisation of the electromagnetic field \cite{Hillery,Vogel,Lewenstein,Drummond,Huttner,Dalton,Dalton2,Dutra,Wubs,Philbin,Zietal} or prefer purely phenomenological approaches to model light scattering through semi-transparent mirrors, like the so-called input-output formalism \cite{CollettGardiner,GardinerCollett,Fan} and different continuous-mode model approaches \cite{Unknown,JMO}. When modelling the transmission of single photons through linear optics networks, we usually employ scattering matrices \cite{Lim,Scheel1}. Unfortunately, the consistency and relationship between these different methods is not yet always well understood \cite{Vogelmore}.

Motivated by our interest in designing novel photonic devices for quantum technology \cite{Kyoseva,Lewis}, this paper discusses the scattering of light through flat mirrors with finite transmission and reflection rates and with the same medium on both sides of the mirror. Here we are especially interested in the case where light approaches the mirror from both sides. As illustrated in Fig.~\ref{fig:intro}, we denote the (real) transmission and reflection rates of light approaching the mirror from the right and from the left by $t_a$ and $r_a$ and by $t_b$ and $r_b$, respectively. Notice that the squares of these rates do not need to add up to one, 
\begin{eqnarray} \label{ineq}
t_s^{ 2} + r_s ^{ 2} \le 1 ~~ {\rm with} ~~ s=a,b \, , 
\end{eqnarray}
meaning that the possible absorption of light in the mirror surface is explicitly taken into account. In other words, we consider mirrors which reduce the amplitudes of incoming wave packets upon transmission and reflection but do not alter them otherwise. 

The field quantisation scheme which we introduce in this paper applies to a wide range of optical mirrors and is strongly motivated by the mirror image method of classical electrodynamics \cite{Jackson}. In the following, we model the electromagnetic field near semi-transparent mirrors by mapping the system onto an equivalent free-space scenario. More concretely, we adopt the same Hamiltonian as in free space and assume that incoming wave packets evolve exactly as they would in the absence of mirrors \cite{EJP}. However, the presence of mirrors changes how and where electric and magnetic field amplitudes are measured. Adopting this point of view, we find that detectors observe superpositions of free-space observables which can be associated with incoming, reflected and transmitted waves. As we shall see below, our approach requires doubling the Hilbert space of the electromagnetic field. 

Although our approach has some similarities with the so-called triplet or normal mode field quantisation schemes of previous authors \cite{Carniglia,Zakowicz,Wu,Eberlein09,Eberlein12,Hammer}, it also provides novel insight into the origin of these modes and extends their potential uses. As we shall see below, the triplet modes which we derive in this paper differ from the triplet modes of Carniglia and Mandel \cite{Carniglia} by phase factors which coincide with the phase factors of the beamsplitter transformations that are routinely used to describe linear optics experiments \cite{Lim,Scheel1}. As a result, our model applies not only to one-sided but also to two-sided semi-transparent mirrors. Moreover, the energy of the mirror surface, i.e.~the energy of mirror images, is explicitly taken into account. As we shall see below, our harmonic oscillator system Hamiltonian $H_{\rm sys}$ can be decomposed into a Hamiltonian $H_{\rm field}$ which describes the energy of the electromagnetic field and a Hamiltonian $H_{\rm mirr}$ which describes the energy of the mirror surface, 
\begin{eqnarray} \label{ineqxxx}
H_{\rm sys} &=& H_{\rm field} + H_{\rm mirr} \, .
\end{eqnarray}
For example, when placing wave packets on only one side of a perfect mirror, we find that half of the energy of the system belongs to the wave packet and the other half belongs to its mirror image. However, when wave packets approach a mirror simultaneously from both sides, then the expectation values of $H_{\rm field}$ and $H_{\rm mirr}$ are in general not the same. 

To test our model, we also calculate the spontaneous decay rate $\Gamma_{\rm mirr}$ and the level shift $\Delta_{\rm mirr}$ of an atom at a fixed distance from a semi-transparent mirror. For simplicity and in order to be consistent with previous authors, we ignore the dependence of the transmission and reflection rates of the mirror on the frequency, polarisation and direction of the incoming light. Doing so, we find that the presence of the mirror alters $\Gamma_{\rm mirr}$ and $\Delta_{\rm mirr}$ as previously predicted for a wide range of situations. As one would expect, we find that a perfect mirror has the same effect as the dipole-dipole interaction between an atom and a mirror atom \cite{dipdip}. When $r_{a} = r_{b} = 1$ and $t_{a} = t_{b}=0$, our approach reproduces the sub- and super-radiance of an atom in front of a perfect mirror which is in good agreement with experimental findings \cite{Drexhage,Eschner,Delsing}. In addition, our calculations cover the case of absorbing mirrors. For example, we find that $\Gamma_{\rm mirr} = \Gamma_{\rm free}$ and $\Delta_{\rm mirr} =0$ when $r_{a} = r_{b} = 0$ and $t_{a} = t_{b}=1$, as it should for an atom in free space. 
 
There are five sections in this paper. Sec.~\ref{CIM} reviews classical electrodynamics and maps the scattering of light on flat surfaces onto analogous free space scenarios. Sec.~\ref{quantum image method}, reviews the properties of the quantised electromagnetic field in free space \cite{EJP} and derives the observables of the quantised electromagnetic field near semi-transparent mirrors. In Sec.~\ref{application}, we test the proposed field quantisation scheme by deriving the master equation of an atom in front of a semi-transparent mirror. Demanding that an atom at a relatively large distance from the mirror surface decays with the same spontaneous decay rate as an atom in free space allows us to determine two previously unknown normalisation factors. Finally, we review our findings in Sec.~\ref{conclusions}. Some more mathematical details can be found in Apps.~\ref{appphases}-\ref{appD}.

\section{Classical light scattering} \label{CIM}

In this section, we review light scattering in classical electrodynamics \cite{Jackson}. We begin by considering light propagation in free space in only one dimension. Subsequently we describe the reflection of light by a one-sided perfect mirror and by a two-sided semi-transparent mirror before considering light scattering in three dimensions. 

\subsection{Free space} \label{WPFreeSpace}

\begin{figure*}[t]
	\centering
	\includegraphics[width=1 \textwidth]{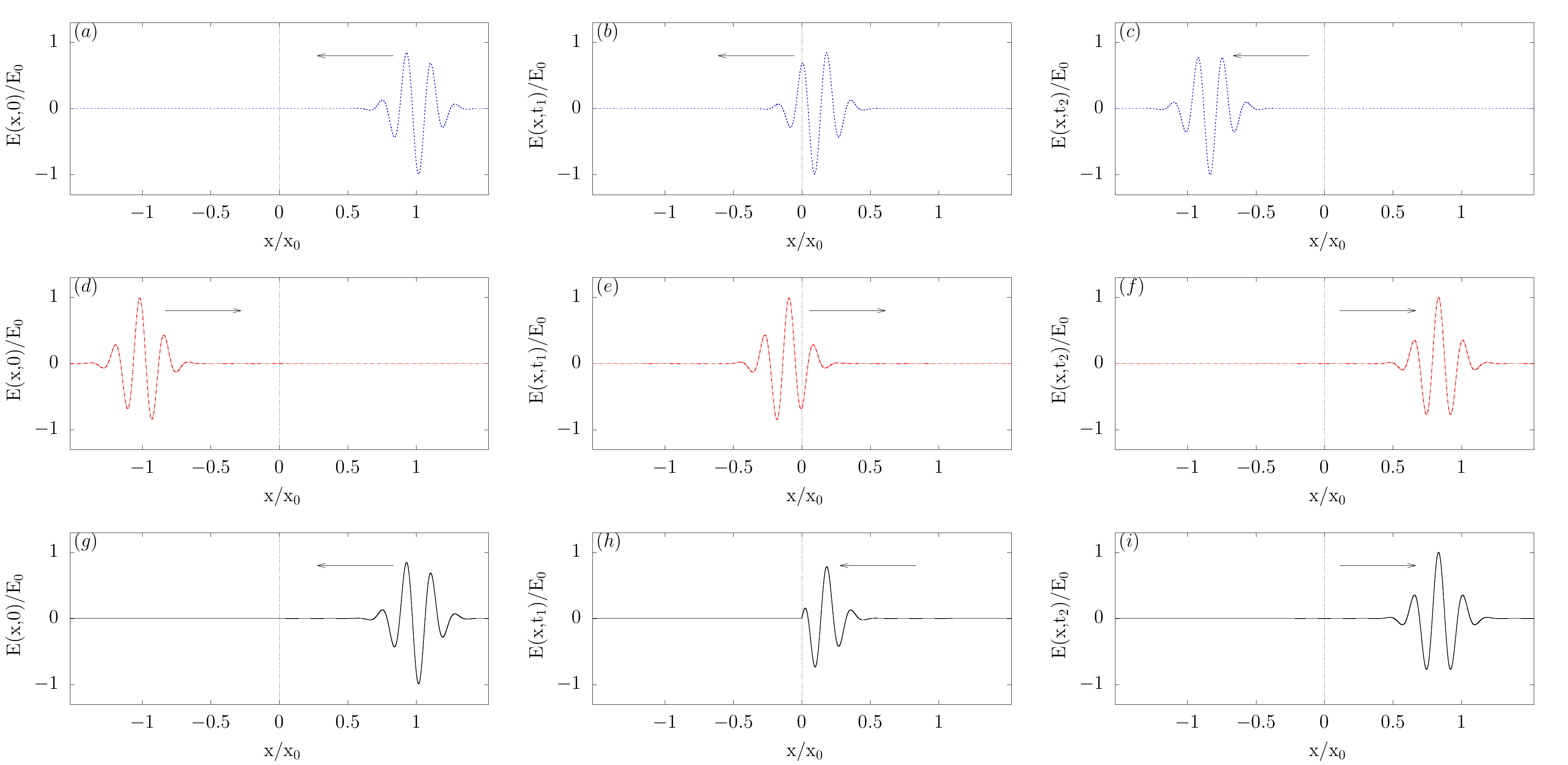}
\caption{[Colour online] Plots (a)--(c) show a left-travelling Gaussian wave packet (blue dotted line) with $E_{\rm free}(x,0) = E_0 \, {\rm e}^{-(x-x_{0})^{2}/2 \sigma^2} \, {\rm e}^{{\rm i}k_{0}x} + {\rm c.c.}$ where $E_0$ and $x_0$ are free parameters and where $k_{0} x_{0} = - 6$, $\sigma = \left( 1/\sqrt{2} \right) x_{0}$, $t_{1} = 0.89 x_{0} /c$ and $t_{2} = 1.83 x_{0}/c$. Moreover, plots (d)--(f)
show a right-travelling Gaussian wave packet (red dotted and dashed line). At $t=0$, the blue wave packet ((a)--(c)) can be interpreted as a real wave packet, while the red wave packet ((d)--(f)) constitutes its mirror image. When the wave packets cross over at $x=0$, the red wave packet becomes real, while the blue one becomes the image. Moreover, plots (g)--(i) show the sum of the red and the blue electric field contribution on the right side of the mirror (black solid line), which evolves like a wave packet approaching a perfectly reflecting mirror.}
	\label{figure:output}
\end{figure*}

In free space, i.e.~in a medium with permittivity $\varepsilon$ and permeability $\mu$ and in the absence of any charges or currents, we describe the dynamics of the electromagnetic field by Maxwell's equations, 
\begin{eqnarray} \label{eq:maxwell}
&& \hspace*{-0.5cm} \nabla \cdot \mathbf{E}_{\rm free} (\mathbf{r},t) = 0 \, , ~~ 
\nabla \cdot \mathbf{B}_{\rm free} (\mathbf{r},t) = 0 \, , \nonumber \\
&& \hspace*{-0.5cm} \nabla \times \mathbf{E}_{\rm free} (\mathbf{r},t) = - \dot{\mathbf B}_{\rm free} (\mathbf{r},t) \, , \nonumber\\
&& \hspace*{-0.5cm} \nabla \times \mathbf{B}_{\rm free} (\mathbf{r},t) = \varepsilon  \mu \, \dot{\mathbf E}_{\rm free} (\mathbf{r},t) \, . 
\end{eqnarray}
Here, $\mathbf{E}_{\rm free} (\mathbf{r},t)$ and $\mathbf{B}_{\rm free} (\mathbf{r},t)$ denote, respectively, the electric and the magnetic field vectors at position $\mathbf{r}$ and at a time $t$. Suppose we are only interested in wave packets with wave vectors ${\bf k} = (k,0,0)$ which propagate along the $x$-axis. In the following, we choose our coordinate system such that $\mathbf{E}_{\rm free}({\bf r},t) = (0,E_{\rm free}(x,t),0)$ and $\mathbf{B}_{\rm free}({\bf r},t) = (0,0,B_{\rm free}(x,t))$ for linear-polarised light with $\lambda = 1$. Moreover, we assume that $\mathbf{E}_{\rm free}({\bf r},t) = (0,0,E_{\rm free}(x,t))$ and $\mathbf{B}_{\rm free}({\bf r},t) = (0,- B_{\rm free}(x,t),0)$ for linear-polarised light with $\lambda = 2$. For these field vectors, Maxwell's equations simplify to 
\begin{eqnarray} \label{Maxwell}
\partial_{x} E_{\rm free}(x,t) &=& - \partial_{t} B_{\rm free}(x,t) \, , \nonumber\\
\partial_{x} B_{\rm free}(x,t) &=& \varepsilon \mu \, \partial_{t} E_{\rm free}(x,t) \, .
\end{eqnarray}
Eliminating the magnetic field from Eq.~\eqref{Maxwell}, we obtain a well-defined one-dimensional wave equation for the electric field,
\begin{eqnarray} \label{eq:waveeqn}
\partial_{x}^{2} E_{\rm free}(x,t) &=& \varepsilon \mu \, \partial_{t}^{2} E_{\rm free}(x,t) \, .
\end{eqnarray}
The general solution $E_{\rm free}(x,t)$ of this equation is a superposition of travelling waves with positive and negative wave numbers $k$ and positive frequencies $\omega$. Analogously, one can show that the general solutions to Maxwell's equations in three dimensions are superpositions of travelling waves with wave vectors $\mathbf{k}$, polarisations $\lambda = 1,2$ and frequencies $\omega$ which obey the following relation 
\begin{eqnarray} \label{eq:Kramerb}
\omega &=& \| {\bf k} \| /\sqrt{\varepsilon \mu} = \| {\bf k} \| \, c \, , 
\end{eqnarray}
where $c$ denotes the speed of light \cite{Jackson}.

\subsection{One-sided perfect mirrors} \label{classicallightscattering}

Now we have a closer look at what happens when a wave packet, which travels along the $x$-axis, approaches a one-sided perfect mirror at $x=0$ from the right. In the presence of the mirror, the electric field $E_{\rm mirr}(x,t)$ still needs to obey Maxwell's equations. In addition, it needs to obey the boundary condition
\begin{eqnarray} \label{eq:boundarycondition}
E_{\rm mirr}(0,t) &=& 0 
\end{eqnarray}
at all times $t$, since the mirror surface charges move freely and are able to immediately compensate for any non-zero electric field contributions. The easiest way of deriving electric and magnetic field solutions in this situation is to apply the mirror image method \cite{Jackson}. This method suggests to write the electric field $E_{\rm mirr}(x,t)$ and its accompanying magnetic field $B_{\rm mirr} (x,t)$ as
\begin{eqnarray} \label{eq:E(x)16}
E_{\rm mirr}(x,t) &=& \left[ E_{\rm free}(x,t) - E_{\rm free}(-x,t) \right] \Theta(x) \, , \nonumber \\
B_{\rm mirr}(x,t) &=& \left[ B_{\rm free}(x,t) + B_{\rm free}(-x,t) \right]  \Theta(x)  ~~
\end{eqnarray}
with the Heaviside step function $\Theta(x)$ defined as
\begin{eqnarray} \label{eq:Heaviside1}
\Theta(x) = \left\{
  \begin{array}{@{}cc@{}}
  1 \, & {\rm for} ~ x \ge 0 \, , \\
  0 \, & {\rm for} ~ x < 0 \, . \\
  \end{array}\right. 
\end{eqnarray}
One can easily check that, at all times, this solution obeys Eq.~\eqref{eq:boundarycondition} and Maxwell's equations, since it is a superposition of free field solutions. Upon reflection, the electric field changes sign, while the magnetic field amplitude remains the same \footnote{Notice that, if $E_{\rm free}(x,t)$ and $B_{\rm free}(x,t)$ are consistent with Maxwell's equations, then the same applies to $E_{\rm free}(-x,t)$ and $- B_{\rm free}(x,t)$.}. 

One way of interpreting Eq.~\eqref{eq:E(x)16} is to say that the mirror produces a mirror image of any incoming wave packet. The mirror image has the same shape as the original wave packet but its components travel with negative electric field amplitudes in the opposite direction. Propagating any incoming wave packet {\em and} its mirror image simultaneously as in free space and adding the respective field amplitudes of both contributions yields exactly the same electric and magnetic fields as Eq.~\eqref{eq:E(x)16} as long as we restrict ourselves to the $x\ge 0$ half space. This is illustrated in Fig.~\ref{figure:output}. Fig.~\ref{figure:output}(a)--(c) and Fig.~\ref{figure:output}(d)--(f) show a left-moving and a right-moving wave packet, respectively, at three different times. The two wave packets cross over the mirror location at $x=0$ at the same time. Adding the electric field contributions on the right side of the mirror, as done in Fig.~\ref{figure:output}(g)--(i), reproduces the dynamics of an incoming wave packet approaching a perfect mirror from the right.  

An alternative way of interpreting Eq.~\eqref{eq:E(x)16} is to say that the mirror introduces a {\em mirror image detector} at a position $-x$ for field amplitude measurements at a position $x$. In addition, we assume that any incoming wave packets propagate exactly as they would in free space. Suppose the image detector for electric field measurements measures $- E_{\rm free}(- x,t)$, while the original detector measures $E_{\rm free}(x,t)$. Moreover, suppose that the total electric field $E_{\rm mirr}(x,t)$ in the presence of the one-sided perfectly-reflecting mirror equals the sum of the amplitudes seen by both the original and the image detector. This approach too reproduces $E_{\rm mirr}(x,t)$ in Eq.~\eqref{eq:E(x)16}. Analogously, we can construct mirror image detectors for magnetic field measurements.

\subsection{Two-sided semi-transparent mirrors} \label{IID}

Next let us have a closer look at what happens when wave packets which travel along the $x$-axis approach a semi-transparent mirror from both sides. As illustrated in Fig.~\ref{fig:intro}, we denote the (real) transmission and reflection rates of the mirror by $t_{a}$, $t_{b}$, $r_{a}$ and $r_{b}$, respectively. Assuming that the mirror only affects the amplitudes but not the shape of incoming wave packets, we can again write the electric field amplitude $E_{\rm mirr}(x,t)$ as a sum of free-space solutions, 
\begin{eqnarray} \label{eq:E(x)24}
E_{\rm mirr}(x,t) &=& E_{\rm free}^{(a)}(x,t) \, \Theta(x) + E_{\rm free}^{(b)}(x,t) \, \Theta(-x) \nonumber \\
&& \hspace*{-1cm} + \left[ r_{a} \, E_{\rm free}^{(a)}(-x,t,\varphi_1) + t_{b} \, E_{\rm free}^{(b)}(x,t,\varphi_2) \right]  \Theta(x) \nonumber \\
&& \hspace*{-1cm} + \left[ r_{b} \, E_{\rm free}^{(b)}(-x,t,\varphi_3) + t_{a} \, E_{\rm free}^{(a)}(x,t,\varphi_4) \right] \Theta(-x) \, , \nonumber \\
\end{eqnarray}
where each term is weighted with its respective rate. One difference to the case of one-sided perfect mirrors is that we now need superscripts to distinguish electric field contributions which originate from different sides of the mirror. In Eq.~(\ref{eq:E(x)24}), the superscripts $(a)$ and $(b)$ are chosen such that, at $t=0$,  
\begin{eqnarray} \label{eq:E(x)22}
E_{\rm free}^{(a)}(x,0) &=& E_{\rm mirr}(x,0) \, \Theta(x) \, , \nonumber \\
E_{\rm free}^{(b)}(x,0) &=& E_{\rm mirr}(x,0) \, \Theta(-x) \, .
\end{eqnarray}
Moreover, $E_{\rm free}^{(s)}(x,t,\varphi) $ is defined such that its amplitude differs from $E_{\rm free}^{(s)}(x,t)$ only by a phase shift $\varphi$. Unfortunately, Eq.~\eqref{eq:E(x)24} applies only for positive times $t$. For $t<0$, the weighting of the individual electric field contributions becomes incorrect. When evolving a wave packet backwards in time, its amplitude should increases whenever it passes through the mirror surface, however, the rates in Eq.~\eqref{eq:E(x)24} are all smaller than unity.

One can easily check that Eq.~\eqref{eq:E(x)24} solves Maxwell's equations, since it is again a superposition of free space solutions. It also produces the expected long-term dynamics for the scattering of incoming wave packets through a two-sided semi-transparent mirror. Moreover, it takes possible relative phase shifts within the mirror surface into account. The electric field amplitude $E_{\rm mirr}(x,t)$ no longer satisfies the boundary condition in Eq.~\eqref{eq:boundarycondition}. Semi-transparent mirrors do not have enough surface charges to compensate all electric field amplitudes. To ensure that maximum interference of the electric field on one side of the mirror implies minimum interference on the other, we assume in the following (c.f.~App.~\ref{appphases} for more details) that
\begin{eqnarray} \label{phases}
\varphi_1 - \varphi_2 + \varphi_3 - \varphi_4 &=& \pm (2n+1) \, \pi \, , 
\end{eqnarray}
where $n$ is an integer. This condition includes lossless semi-transparent mirrors \cite{Lewis1,Lewis2}. Moreover, Eq.~\eqref{eq:E(x)24} includes free space as a limiting case which corresponds to $r_a=r_b=0$, $t_a = t_b=1$ and $\varphi_2=\varphi_4=0$. In addition, Eq.~\eqref{eq:E(x)24} reproduces the one-sided perfect mirror case (c.f.~Eq.~\eqref{eq:E(x)16}), if we choose $r_a=r_b=1$, $t_a = t_b=0$ and $\varphi_1=\varphi_3=\pi$. In general, three of the phase factors $\varphi_i$ depend on the properties of the mirror surface but can be determined relatively easily experimentally. The remaining fourth parameter is established when the interference of wave packets originating from different sides of the mirror is first observed.

As pointed out already in Sec.~\ref{Intro}, the possible absorption of light in the mirror surface is explicitly taken into account in this paper, thereby allowing a portion of the energy of incoming wave packets to be dissipated within the mirror surface. However, we assume here that absorption does not affect the shape of the incoming wave packets. It only reduces their amplitudes. Moreover, we assume that the reflection and transmission rates of the mirror do not depend on the frequency of the incoming light. For simplicity, we also ignore the existence of evanescent wave solutions of Maxwell's equations, i.e.~we only consider the electromagnetic field at a certain minimum distance away from the mirror surface \cite{Carniglia}.

\subsection{Generalisation to semi-transparent mirrors in three dimensions}

\begin{figure}[t] 
	\centering
        \includegraphics[width=0.5\textwidth]{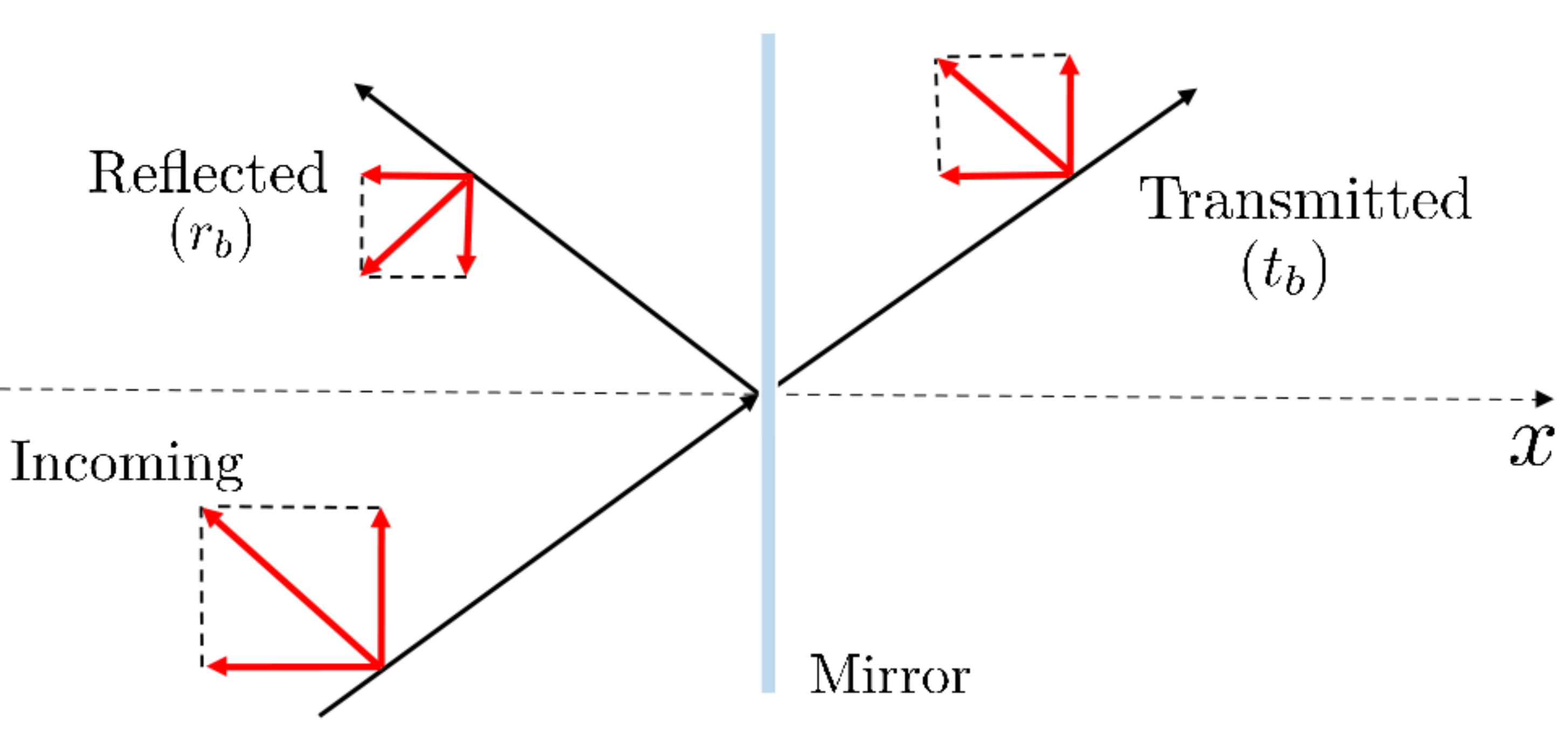} 	
\caption{[Colour online] Schematic view of a semi-transparent mirror with light incident from the left. The solid black lines incident on the mirror indicate the direction of the wave vector of the incoming light. To predict the effect of the mirror, we decompose the electric field amplitudes of incoming wave packets into its parallel and perpendicular components (red lines) with respect to the mirror surface. As illustrated, transmission and reflection reduces these components by a factor which equals the corresponding transmission and reflection rate. Notice that these rates have to be the same for parallel and perpendicular field amplitudes. Otherwise, the electric field vector would not remain orthogonal to the corresponding wave vector ${\bf k}$. However, parallel electric field components obtain a minus sign upon reflection due to the rearrangement of mirror surface charges.}
	\label{fig:intro2}
\end{figure}

Finally, we analyse the dynamics of the electromagnetic field near a semi-transparent mirror with light approaching the mirror at any possible angle (c.f.~Fig.~\ref{fig:intro2}). Again we assume that the mirror does not affect the dynamics of incoming wave packets but changes how and where the electric and magnetic field amplitudes ${\bf E}_{\rm mirr}({\bf r},t)$ and ${\bf B}_{\rm mirr}({\bf r},t)$ are detected. Suppose an electric field detector observes incoming and transmitted wave packets at a position ${\bf r} = (x,y,z)$. In addition, the detector sees the electric field amplitudes of reflected wave packets which equal the electric field amplitudes of freely-propagating wave packets at a position $\widetilde{\bf r} $ with
\begin{eqnarray}  \label{eq:38b}
\widetilde{\bf r} &=& (-x,y,z) \, .
\end{eqnarray}
Here the coordinate system is chosen such that the mirror lies in the $x=0$ plane.

For perfectly-reflecting mirrors, strict boundary conditions need to be imposed. The transversal component of electric field amplitudes and the normal components of magnetic field amplitudes both need to constantly vanish on the mirror surface \cite{Jackson}. As a result, the $x$ component of the electric field needs to remains unchanged upon reflection, while its $y$ and $z$ components acquire a minus sign. In the following, we assume that the same applies for the electric field amplitude of wave packets which have been reflected by a semi-transparent mirror, as illustrated in Fig.~\ref{fig:intro2}. In general, a semi-transparent mirror causes phase shifts of electric field amplitudes upon transmission and reflection. Moreover, these field amplitudes need to be multiplied by transmission and reflection rates. Taking this into account, we now find that the electric field ${\bf E}_{\rm mirr}({\bf r},t)$ is the sum of six  contributions,
\begin{eqnarray} \label{eq:E(x)25}
{\bf E}_{\rm mirr}({\bf r},t) &=& {\bf E}_{\rm free}^{(a)}({\bf r},t) \, \Theta(x) + {\bf E}_{\rm free}^{(b)}({\bf r},t) \, \Theta(-x) \nonumber \\
&& \hspace*{-0.7cm} + \left[ r_{a} \, \widetilde {\bf E}_{\rm free}^{(a)}(\widetilde{\bf r},t,\varphi_1) + t_{b} \, {\bf E}_{\rm free}^{(b)}({\bf r},t, \varphi_2) \right]  \Theta(x) \nonumber \\
&& \hspace*{-0.7cm} + \left[ r_b \, \widetilde {\bf E}_{\rm free}^{(b)}(\widetilde{\bf r},t,\varphi_3) + t_a \, {\bf E}_{\rm free}^{(a)}({\bf r},t,\varphi_4) \right] \Theta(-x) \, , \nonumber \\
\end{eqnarray}
where $x$ refers to the $x$-component of ${\bf r}$. Here ${\bf E}_{\rm free}^{(s)}({\bf r},t)$ denotes an electric field  free-space solution of Maxwell's equations. Moreover, $\widetilde {\bf E}_{\rm free}^{(s)}({\bf r},t)$ is defined such that it differs from ${\bf E}_{\rm free}^{(s)}({\bf r},t)$ only by the sign of its $x$-component. The superscripts $(a)$ and $(b)$ help again to distinguish initial electric field contributions on the left and on the ride side of the mirror. The phase factors $\varphi_i$ play the same role as the phase factors 
$\varphi_i$ in Eq.~(\ref{eq:E(x)24}). The same transmission and reflection rates need to apply to vertical and horizontal electric field components. Otherwise, ${\bf E}_{\rm mirr}({\bf r},t) $ is not orthogonal to the corresponding wave vector ${\bf k}$.

\section{A quantum image detector method to light scattering} \label{quantum image method}

Sec.~\ref{CIM} maps the scattering of light through a two-sided semi-transparent mirror onto an analogous free-space scenario with detectors and mirror image detectors. In this section, we use the analogy of both situations to quantise the electromagnetic field near a two-sided semi-transparent mirror. In the following, we derive its system Hamiltonian $H_{\rm sys}$ and the electric field observable ${\bf E}_{\rm mirr}({\bf r})$ as a function of the transmission and reflection rates of the mirror.

\subsection{Free space} \label{IIIA}

To quantise the electromagnetic field in free space in one dimension, we first notice that the field Hamiltonian $H_{\rm sys}$ can be deduced from experimental observations \cite{EJP}. These confirm that light propagating along the $x$-axis consists of basic energy quanta (photons) with positive and negative wave numbers $k$. Moreover, we know that the energy of a single photon in the $k$-mode equals $\hbar \omega$ with $\omega$ given in Eq.~\eqref{eq:Kramerb}. This implies that the electromagnetic field is a collection of harmonic oscillator modes. Taking this into account, only considering light with polarisation $\lambda =1$ which propagates along the $x$-axis and introducing the bosonic annihilation and creation operators $a_k$ and $a_k^{\dagger}$ with 
\begin{eqnarray} \label{eq:bosoniccommutationrelation}
\big[ a_k , a_{k'}^{\dagger} \big] &=& \delta(k - k') \, ,
\end{eqnarray}
the electromagnetic field Hamiltonian $H_{\rm sys}$ can be written as 
\begin{eqnarray} \label{eq:Hfree1}
H_{\rm sys} &=& \int_{- \infty}^{\infty} {\rm d} k \, \hbar \omega \, a_k^{\dagger} a_k \, .
\end{eqnarray}
Moreover, we know that the observable for the energy stored inside the electromagnetic field $H_{\rm field}$ equals
\begin{eqnarray} \label{eq:Hfree12}
H_{\rm field} &=& {A \over 2} \int_{- \infty}^{\infty} {\rm d} x \, \left[ \varepsilon \, E_{\rm free}(x)^2 + {1 \over \mu} \, B_{\rm free}(x)^2 \right] , ~~~
\end{eqnarray}
where $A$ denotes the area in the $y$-$z$ plane in which $H_{\rm sys}$ and $H_{\rm field}$ are defined. Here $E_{\rm free}(x)$ and $B_{\rm free}(x)$ denote the free-space observables of the electric and the magnetic field amplitudes, respectively. In the absence of any mirrors, both Hamiltonians $H_{\rm sys} $ and $H_{\rm field}$ coincide with each other up to a constant which is known as zero-point energy. 

Comparing Eqs.~\eqref{eq:Hfree1} and \eqref{eq:Hfree12} suggests that the electric field observable $E_{\rm free}(x)$ and the magnetic field observable $B_{\rm free}(x)$ are linear superpositions of annihilation and creation operators. One can calculate the corresponding coefficients demanding that expectation values of $E_{\rm free}(x)$ and $B_{\rm free}(x)$ evolve according to Maxwell's equations. Doing so and normalising $E_{\rm free}(x)$ and $B_{\rm free}(x)$ such that $H_{\rm sys} = H_{\rm field}$ yields \cite{EJP} 
\begin{eqnarray} \label{eq:1DEField}
E_{\rm free}(x) &=& {{\rm i} \over 2} \, \int_{-\infty}^{\infty} {\rm d} k \, \sqrt{\hbar \omega \over \pi \varepsilon A} \, {\rm e}^{ {\rm i}kx} \, a_k + {\rm H.c.} \, , \nonumber \\
B_{\rm free}(x) &=& - {{\rm i} \over 2c} \, \int_{-\infty}^{\infty} {\rm d} k \, \sqrt{\hbar \omega \over \pi \varepsilon A} \, {\rm e}^{ {\rm i}kx} \, a_k \, \rm{sign}(k) ~~ \nonumber \\
&& + {\rm H.c.}
\end{eqnarray}
Analogously, one can derive the electric and magnetic field observables for $\lambda=2$ polarised light. These are of the same form as $E_{\rm free}(x)$ and $B_{\rm free}(x)$ in Eq.~\eqref{eq:1DEField} up to an overall minus sign of the magnetic field.

\subsection{One-sided perfect mirrors} \label{finsec}

Next, we consider again a one-sided perfect mirror in the $x=0$ plane with non-zero field components only on its right side. From experience we know that a photon of frequency $\omega$ has the energy $\hbar \omega$, even in the presence of a mirror. Hence, when using the same notion of photons as in free space, we find that the system Hamiltonian $H_{\rm sys}$ in the presence of a perfect mirror must be the same as the free space Hamiltonian $H_{\rm sys}$ in Eq.~\eqref{eq:Hfree1}. Moreover, as we have seen in Sec.~\ref{CIM}, wave packets evolve essentially as in free space, even in the presence of mirrors. What changes is how and where electromagnetic field amplitudes are measured. These are now the sum of the field amplitudes seen by original detectors and the field amplitudes seen by mirror image detectors in the corresponding free-space scenario. Taking this into account, Eq.~\eqref{eq:E(x)16} suggests that 
\begin{eqnarray} \label{eq:E(x)31}
E_{\rm mirr}(x) &=& {1 \over \eta} \left[ E_{\rm free}(x) - E_{\rm free}(-x) \right]  \Theta(x) \, , \nonumber \\
B_{\rm mirr}(x) &=& {1 \over \eta} \left[ B_{\rm free}(x) + B_{\rm free}(-x) \right]  \Theta(x) \, ,
\end{eqnarray} 
with $E_{\rm free}(x)$ and $B_{\rm free}(x)$ given in Eq.~\eqref{eq:1DEField} and with $\eta$ denoting a normalisation factor.

As we shall see below in Sec.~\ref{application}, assuming that an atom far away from the mirror surface decays exactly as it would in free space, implies
\begin{eqnarray} \label{final6}
\eta &=& \sqrt{2} \, .
\end{eqnarray}
Taking this into account, introducing standing-wave photon annihilation operators $\xi_k$,
\begin{eqnarray} \label{eq:E(x)32c}
\xi_k = {1 \over \sqrt{2}}  \left(a_k - a_{-k} \right) ~~& {\rm with} &~~ \xi_{-k} = - \xi_k 
\end{eqnarray}
and combining Eqs.~(\ref{eq:1DEField}) and \eqref{eq:E(x)31}, the field operators $E_{\rm mirr}(x)$ and $B_{\rm mirr}(x)$ simplify to 
\begin{eqnarray} \label{eq:E(x)32b}
E_{\rm mirr}(x) &=& {{\rm i} \over 2} \, \int_{-\infty}^{\infty} {\rm d} k \, \sqrt{\hbar \omega \over \pi \varepsilon A} \, {\rm e}^{ {\rm i}kx} \, \xi_k \, \Theta(x) + {\rm H.c.} \, , \nonumber \\
B_{\rm mirr}(x) &=& - {{\rm i} \over 2c} \, \int_{-\infty}^{\infty} {\rm d} k \, \sqrt{\hbar \omega \over \pi \varepsilon A} \, {\rm e}^{ {\rm i}kx} \, \xi_k \, {\rm sign}(k) \, \Theta(x) ~~ \nonumber \\
&& + {\rm H.c.}
\end{eqnarray}
Moreover, we know that the energy of the electromagnetic field on the right side of the mirror equals 
\begin{eqnarray} \label{eq:Hfree12b}
H_{\rm field} &=& {A \over 2} \int_0^{\infty} {\rm d} x \, \left[ \varepsilon \, E_{\rm mirr}(x)^2 + {1 \over \mu} \, B_{\rm mirr}(x)^2 \right] . ~~~~~
\end{eqnarray}
Proceeding as described in App.~\ref{AppendixA2}, we therefore find that 
\begin{eqnarray} \label{final}
H_{\rm field} &=& \int_0^{\infty} {\rm d} k \, \hbar \omega \, \xi_{k}^{\dagger} \xi_{k}   
\end{eqnarray}
up to a constant. One can easily check that this field Hamiltonian commutes with $H_{\rm sys}$ and that its expectation values are conserved.

However, notice that $H_{\rm sys}$ and $H_{\rm field}$ are no longer the same. For example, suppose a wave packet approaches the mirror from the right. In this case, exactly half of the population of the electromagnetic field is in the antisymmetric $\xi_k$ modes. All other population is in orthogonal (symmetric) modes and
\begin{eqnarray} 
\langle H_{\rm field} \rangle &=& {1 \over 2} \langle H_{\rm sys} \rangle \, .
\end{eqnarray}
Only half of the energy of the system is stored in the electromagnetic field in this case. The other half belongs to the mirror image of the incoming wave packet. As illustrated in Fig.~\ref{figure:output}, our mirror scenario is indeed equivalent to having two wave packets travelling in opposite directions in free space. The difference between $H_{\rm sys}$ and $H_{\rm field}$ is the observable $H_{\rm mirr}$ for the energy of the mirror surface charges (c.f.~Eq.~\eqref{ineqxxx}).

Previous quantisation schemes for the electromagnetic field in front of a perfect mirror ignore the energy of the mirror surface (see e.g.~Refs.~\cite{Morawitz,Meschede,Drabe,Matloob,Dorner,Beige}). Nevertheless, they are consistent with our approach. Suppose a wave packet approaches a one-sided perfect mirror from the right, as illustrated in Fig.~\ref{figure:output}(a)--(c). If we are only interested in the electromagnetic field on the right side of the mirror, we can extend the initial state to the left. When doing so, we introduce the mirror image shown in Fig.~\ref{figure:output}(d)--(f) which is equivalent to having an initial state with population only in the $\xi_k$ modes. For these modes, the field observables $H_{\rm sys}$, $E_{\rm mirr}(x)$ and $B_{\rm mirr}(x)$ are exactly the same as in free space. However, in this paper, we do not restrict ourselves to a subset of possible initial states and allow mirrors to be approached by arbitrary wave packets from both sides. 

\subsection{Two-sided semi-transparent mirrors} \label{semisec}

As shown in Sec.~\ref{IID}, the dynamics of wave packets which approach a semi-transparent mirror from both sides depends on whether they originate from the left or from the right side. To be able to distinguish both cases, we now replace the Hilbert space ${\cal H}$ for the modelling of the electromagnetic field in free space by the tensor product of two free-space Hilbert spaces $\mathcal{H}^{(a)}$ and $\mathcal{H}^{(b)}$,
\begin{eqnarray} \label{eq:Hilbertspace}
\mathcal{H} &\to & \mathcal{H}^{(a)} \otimes \mathcal{H}^{(b)} \, . 
\end{eqnarray}
Considering again only light travelling along the $x$-axis and denoting the corresponding photon annihilation operators belonging to $\mathcal{H}^{(a)}$ and $\mathcal{H}^{(b)}$ by $a_k$ and $b_k$, respectively, the system Hamiltonian $H_{\rm sys}$ of the mirror surfaces and the surrounding electromagnetic fields becomes
\begin{eqnarray} \label{eq:Hfree23}
H_{\rm sys} &=& \int_{- \infty}^{\infty} {\rm d}k \, \hbar \omega \left[ a_k^{\dagger} a_k + b_k^{\dagger} b_k \right] \, .
\end{eqnarray}
As in Sec.~\ref{IID}, the superscripts $(a)$ and $(b)$ indicate quantum states which originate from the right and the left half space of the mirror, respectively. 

Moreover, Eq.~\eqref{eq:E(x)24} suggests that the observable $E_{\rm mirr}(x)$ of the electric field near a semi-transparent mirror is a superposition of free-space observables,
\begin{eqnarray} \label{eq:E(x)43c}
E_{\rm mirr}(x) &=& {1 \over \eta_a} E_{\rm free}^{(a)}(x) \, \Theta (x) + {1 \over \eta_b} E_{\rm free}^{(b)}(x) \, \Theta (-x) \nonumber \\
&& \hspace*{-0.5cm} + \left[ {r_{a} \over \eta_a} E_{\rm free}^{(a)}(-x,\varphi_1) + {t_{b} \over \eta_b} E_{\rm free}^{(b)}(x,\varphi_2) \right] \Theta (x) \nonumber \\
&& \hspace*{-0.5cm} + \left[ {r_{b} \over \eta_b} E_{\rm free}^{(b)}(-x,\varphi_3) + {t_{a} \over \eta_a} E_{\rm free}^{(a)}(x,\varphi_4) \right] \Theta(-x) \, . \nonumber \\
\end{eqnarray}
The additional argument in $E_{\rm free}^{(s)}(x,\varphi)$ indicates a $\varphi$ phase shift of the electric amplitude with respect to the field amplitude of $E_{\rm free}^{(s)}(x)$. Moreover, the constants $\eta_a$ and $\eta_b$ are normalisation factors. To determine them we need to specify not only transmission and reflection rates but also the medium on either side of the semi-transparent mirror. In Sec.~\ref{application} this is done by calculating the spontaneous decay rate $\Gamma_{\rm mirr}$ of an atom in front of the mirror. Demanding, for example,  that $\Gamma_{\rm mirr}$ simplifies to $\Gamma_{\rm free}$ in Eq.~\eqref{Gamma0} for relatively large atom-mirror distances, as it should in free space, ultimately fixes $\eta_a$ and $\eta_b$ (c.f.~Eq.~\eqref{final10}). 

As in classical physics, the expectation value of $E_{\rm mirr}(x)$ no longer always vanishes at $x=0$. Moreover, the transmission of light through a semi-transparent mirror surface can results in the loss of energy from the electromagnetic field. As pointed out in Sec.~\ref{CIM}, the possible absorption of light is already built into our model. Only the energy of the electromagnetic field and the mirror surface, i.e.~the expectation value of the system Hamiltonian $H_{\rm sys}$ in Eq.~\eqref{eq:Hfree23}, is conserved. However, the expectation value of the electromagnetic field Hamiltonian $H_{\rm field}$ can change in time. In general, there is a continuous exchange of energy between the electromagnetic field and the mirror surface. For example, suppose a wave packet approaches the mirror from the right. After a sufficiently long time, this wave packet turns into two wave packets: one on the left and one on the right side of the mirror. This implies a reduction of the energy stored inside the electromagnetic field by a factor which can be smaller than one (c.f.~Eq.~\eqref{ineq}).

\subsection{Generalisation to semi-transparent mirrors in three dimensions}

Finally, we have again a closer look at the quantised electromagnetic field in three dimensions. In free space, the electric field observable ${\bf E}_{\rm free} ({\bf r})$ at a position ${\bf r}$ for light propagation in three dimensions equals \cite{EJP} 
\begin{eqnarray} \label{eq:36}
{\bf E}_{\rm free}({\bf r}) &=& {{\rm i} \over 4 \pi } \, \sum_{\lambda = 1,2} \int_{\mathbb{R}^3} {\rm d}^3 {\bf k} \, \sqrt{\hbar \omega \over \pi \varepsilon} \, {\rm e}^{{\rm i} {\bf k} \cdot {\bf r}} \, \hat {\bf e}_{{\bf k} \lambda}  \, a_{{\bf k} \lambda} + {\rm H.c.} \nonumber \\
\end{eqnarray}
which sums over all possible photon modes with wave vectors ${\bf k}$ and polarisations $\lambda$. Moreover  $a_{{\bf k}\lambda}$ is the photon annihilation operator of the $({\bf k}, \lambda)$ mode with the bosonic commutator relation
\begin{eqnarray} \label{eq:35}
\left[ a_{{\bf k} \lambda}, a_{{\bf k}' \lambda'}^{\dagger} \right] &=& \delta_{\lambda \lambda'} \, \delta^{3}({\bf k}-{\bf k}') \, .
\end{eqnarray}
The normalised polarisation vectors $\hat {\bf e}_{{\bf k} \lambda} $ in Eq.~\eqref{eq:36} are pairwise orthogonal and $\hat {\bf e}_{{\bf k} \lambda} \cdot {\bf k} = 0$ for all ${\bf k}$. The frequency $\omega$ can be found in Eq.~\eqref{eq:Kramerb} and the constant $\varepsilon$ is the same as in Sec.~\ref{WPFreeSpace}. Moreover, the Hamiltonian $H_{\rm sys}$ of the electromagnetic field in free space in three dimensions equals~\cite{EJP} 
\begin{eqnarray} \label{eq:Hfree22}
H_{\rm sys} &=& \sum_{\lambda = 1,2} \int_{\mathbb{R}^{3}} {\rm d}^3 {\bf k} \, \hbar \omega \, a_{{\bf k} \lambda}^{\dagger} a_{{\bf k} \lambda}
\end{eqnarray}
in analogy to Eq.~\eqref{eq:Hfree1}.

To model a two-sided semi-transparent mirror in the $x=0$ plane, we double again the Hilbert space compared to the above described free-space scenario. Denoting the corresponding photon annihilation operators by $a_{{\bf k} \lambda}$ and $b_{{\bf k} \lambda}$, respectively, the system Hamiltonian $H_{\rm  sys}$ of the electromagnetic field and the mirror surface equals 
\begin{eqnarray} \label{eq:Hfree2222}
H_{\rm sys} &=& \sum_{\lambda = 1,2} \int_{\mathbb{R}^{3}} {\rm d}^3 {\bf k} \, \hbar \omega \, \left[ a_{{\bf k} \lambda}^{\dagger} a_{{\bf k} \lambda} + b_{{\bf k} \lambda}^{\dagger} b_{{\bf k} \lambda} \right] ~~
\end{eqnarray}
in analogy to Eq.~\eqref{eq:Hfree23}. To obtain the observable ${\bf E}_{\rm mirr}({\bf r})$ of the electric field at position ${\bf r}$, we use again the above introduced quantum image detector method. As in Sec.~\ref{CIM}, we assume that wave packets evolve as in free space but that an electric field detector at position ${\bf r}$ observes electric field contributions of incoming, transmitted and reflected wave packets. Reflection changes the sign of the $y$-and the $z$-component of the electric field of incoming wave packets, while their $x$-component remains unaffected. Hence 
\begin{eqnarray} \label{eq:38full}
{\bf E}_{\rm mirr}({\bf r}) &=& {1 \over \eta_a} \, {\bf E}_{\rm free}^{(a)}({\bf r}) \, \Theta(x) + {1 \over \eta_b} \, {\bf E}_{\rm free}^{(b)}({\bf r}) \, \Theta(-x)\nonumber \\
&& + \left[ {r_{a} \over \eta_a} \, \widetilde {\bf E}_{\rm free}^{(a)}(\widetilde{\bf r},\varphi_1) + {t_{b} \over \eta_b} \, {\bf E}_{\rm free}^{(b)}({\bf r},\varphi_2) \right] \Theta(x) \nonumber \\
&& + \left[ {r_{b} \over \eta_b} \, \widetilde {\bf E}_{\rm free}^{(b)}(\widetilde{\bf r},\varphi_3) + {t_{a} \over \eta_a} \, {\bf E}_{\rm free}^{(a)}({\bf r},\varphi_4) \right] \Theta(-x) \nonumber \\
\end{eqnarray}
in analogy to Eq.~(\ref{eq:E(x)25}). The definition of $\widetilde{\bf r}$ can be found in Eq.~(\ref{eq:38b}) and $\widetilde {\bf E}_{\rm free}^{(s)}({\bf r})$ differs from ${\bf E}_{\rm free}^{(s)}({\bf r})$ only by the sign of its $x$-component. The argument in $ \Theta(x)$ refers again to the $x$-component of ${\bf r}$ and the factors $\eta_a$ and $\eta_b$ are again normalisation factors. Similar to Eq.~\eqref{eq:E(x)25}, Eq.~\eqref{eq:38full} contains the perfect mirror limiting case when we set $r_{a} = r_{b} = 1$ and $t_{a}=t_{b}=0$ with $\varphi_{1}=\varphi_{3}=\pi$.
 
\section{The master equation of an atom near a semi-transparent mirror} \label{application}

In this section, we test our model by deriving the spontaneous decay rate and the level shift of an atom at a certain distance $x$ away from the mirror. As we shall see below, in limiting cases like perfect mirror reflection, our results are consistent with the results of other authors \cite{Morawitz,Stehle,Milonni,Kleppner,Sipe,Arnoldus1,Meschede,Drabe,Matloob,Dorner,Beige} and experimental findings \cite{Drexhage,Eschner,Delsing}. In the following, we assume that the atom-mirror distance does not become so large that delay terms have to be taken into account \cite{Dorner}. In addition, it should not be too short in order to avoid the interaction of the system with evanescent field modes. In the limit of relatively large atom-mirror distances $x$, we demand that the spontaneous decay rate $\Gamma_{\rm mirr}$ of the atom coincides with its free-space decay rate $\Gamma_{\rm free}$ in Eq.~\eqref{Gamma0}. Imposing this as a condition allows us to calculate the normalisation factors $\eta_a$ and $\eta_b$ in Eq.~\eqref{eq:38full} as a function of the reflection and transmission rates of the two-sided semi-transparent mirror. The only other assumptions made in this section are standard quantum optical approximations and approximations which are also made by other authors.

\subsection{General derivation}

To obtain the master equations of an atom in front of a semi-transparent mirror, we now follow the approach of Refs.~\cite{He1,Stokes}. Our starting point is the Hamiltonian $H$, which describes the energy of the atom, the free radiation field, the mirror surface and their respective interactions,  
\begin{eqnarray} \label{eq:Ham1}
H &=& H_{\rm atom} + H_{\rm sys} + H_{\rm int} \, .
\end{eqnarray}
Suppose $|1 \rangle$ denotes the ground state of the atom and $|2 \rangle$ is its excited state with energy $\hbar \omega_0$. Then 
\begin{eqnarray} \label{eq:Hdef}
H_{\rm atom} &=& \hbar \omega_0 \, |2 \rangle \langle 2| \, .
\end{eqnarray}
The system Hamiltonian $H_{\rm sys}$ of the electromagnetic field and the semi-transparent mirror can be found in Eq.~\eqref{eq:Hfree2222}. Moreover, the atom-field interaction Hamiltonian $H_{\rm int}$ equals
\begin{eqnarray} \label{eq:Hdef2}
H_{\rm int} &=& e \, \mathbf{D} \cdot \mathbf{E}_{\rm mirr} (\mathbf{r}) 
\end{eqnarray}
in the usual dipole approximation. Here $e$ is the charge of a single electron, $\mathbf{E}_{\rm mirr} (\mathbf{r})$ is the observable of the electric field at the position ${\bf r}$ of the atom, while 
\begin{eqnarray} \label{eq:Hdef3}
\mathbf{D} &=& \mathbf{D}_{12} \, \sigma^{-} + \mathbf{D}^{\star}_{12} \, \sigma^{+} 
\end{eqnarray}
is the atomic dipole moment. Moreover, $\mathbf{D}_{12}$ is a complex vector and $\sigma^{+} = |2 \rangle \langle 1|$ and $\sigma^{-} = |1 \rangle \langle 2|$ are the atomic raising and lowering operators. 

When going into the interaction picture with respect to the free Hamiltonian $H_{0} = H_{\rm atom} + H_{\rm sys}$, the Hamiltonian $H$ in Eq.~(\ref{eq:Ham1}) transforms into the interaction Hamiltonian 
\begin{eqnarray} \label{eq:Ham33}
H_{\rm I}(t) &=& U_0^{\dagger}(t,0) \, H_{\rm int} \, U_0(t,0) \, .
\end{eqnarray}
Combining Eqs.~\eqref{eq:Ham1}--\eqref{eq:38full} and applying the rotating wave approximation yields the interaction Hamiltonian 
\begin{eqnarray} \label{eq:E(x)45}
H_{\rm I}(t) &=& {{\rm i} e \over 4 \pi} \, \sum_{\lambda = 1,2} \int_{\mathbb{R}^3} {\rm d}^3 {\bf k} \, \sqrt{\hbar \omega \over \pi \varepsilon} \, {\rm e}^{- {\rm i} (\omega - \omega_0) t} \nonumber \\
&& \times \Bigg[ \left( \mathbf{D}^{\star}_{12} \cdot  \hat {\bf e}_{{\bf k} \lambda} \right) {\rm e}^{{\rm i} {\bf k} \cdot {\bf r}} \left( {1 \over \eta_a} \, a_{{\bf k} \lambda} + {t_b \over \eta_b} \, b_{{\bf k} \lambda} \right) \nonumber \\
&& - \left( \widetilde{\mathbf{D}}^{\star}_{12} \cdot  \hat {\bf e}_{{\bf k} \lambda} \right) {\rm e}^{{\rm i} {\bf k} \cdot \widetilde{\bf r}} \, {r_{a} \over \eta_a} \, a_{{\bf k} \lambda} \Bigg] \, \sigma^{+} + {\rm H.c.} \, , ~~~~
\end{eqnarray}
if we place the atom on the right side of the mirror, where its $x$-coordinate is positive. Here $\widetilde{\mathbf{D}}_{12} $ differs from ${\mathbf{D}}_{12} $ only by the sign of its $x$-component. Moreover, we chose
\begin{eqnarray}
\varphi_1 = \pi ~~& {\rm and} &~~ \varphi_2 = 0 \, .  
\end{eqnarray}
In this way, our model contains the free-space and the one-sided perfect mirror scenario as limiting cases. However, in general, $\varphi_1$ and $\varphi_2$ might be different from the above choice and need to be determined using classical interference experiments.

Proceeding as described for example in Refs.~\cite{He1,Stokes}, we find that the interaction picture density matrix $\rho_{\rm AI} (t)$ of an atom with spontaneous photon emission evolves according to a master equation of the form
\begin{eqnarray} \label{eq:rho3}
\dot{\rho}_{\rm AI} (t) &=& - {{\rm i} \over \hbar} \left[ H_{\rm cond} \, \rho_{\rm AI} (t) - \rho_{\rm AI} (t) \, H_{\rm cond}^\dagger \right] + {\cal L} (\rho_{\rm AI} (t)) \nonumber \\
\end{eqnarray}
with the conditional Hamiltonian $H_{\rm cond}$ and with the reset operator ${\cal L} (\rho_{\rm AI} (t))$ given by
\begin{eqnarray} \label{eq:rho4}
H_{\rm cond} &=& \frac{1}{\Delta t} \int \limits_{t}^{t + \Delta t} {\rm d}t^{\prime} \, \langle 0| H_{\rm I} (t') |0 \rangle \nonumber \\
&& - \frac{{\rm i}}{\hbar \Delta t} \int \limits_{t}^{t + \Delta t} {\rm d}t^{\prime} \int \limits_{t}^{t^{\prime}} {\rm d}t^{\prime \prime} \,\langle 0| H_{\rm I} (t') H_{\rm I} (t'') |0 \rangle ~~~~
\end{eqnarray}
and
\begin{eqnarray} \label{eq:rho4xxx}
{\cal L} (\rho_{\rm AI} (t)) &=& \frac{1}{\hbar^{2} \Delta t} \int_{t}^{t + \Delta t} {\rm d}t' \int_{t}^{t + \Delta t} {\rm d}t'' \nonumber \\
&& \times {\rm Tr}_{\rm field} \Big( \, H_{\rm I}(t') |0 \rangle \, \rho_{\rm AI}(t) \, \langle 0| H_{\rm I}(t'') \, \Big) . ~~~~~
\end{eqnarray}
In the case of an environment that monitors the spontaneous emission of photons, the non-Hermitian Hamiltonian $H_{\rm cond}$ describes the time evolution of the atom under the condition of no photon emission, while the reset operator ${\cal L} (\rho_{\rm AI} (t))$ denotes the un-normalised state of the atom in the case of an emission at $t$ \cite{He1}.

\subsection{Atomic master equations} \label{results}

\begin{figure*}[t]
    \centering
		\includegraphics[width=0.98\textwidth]{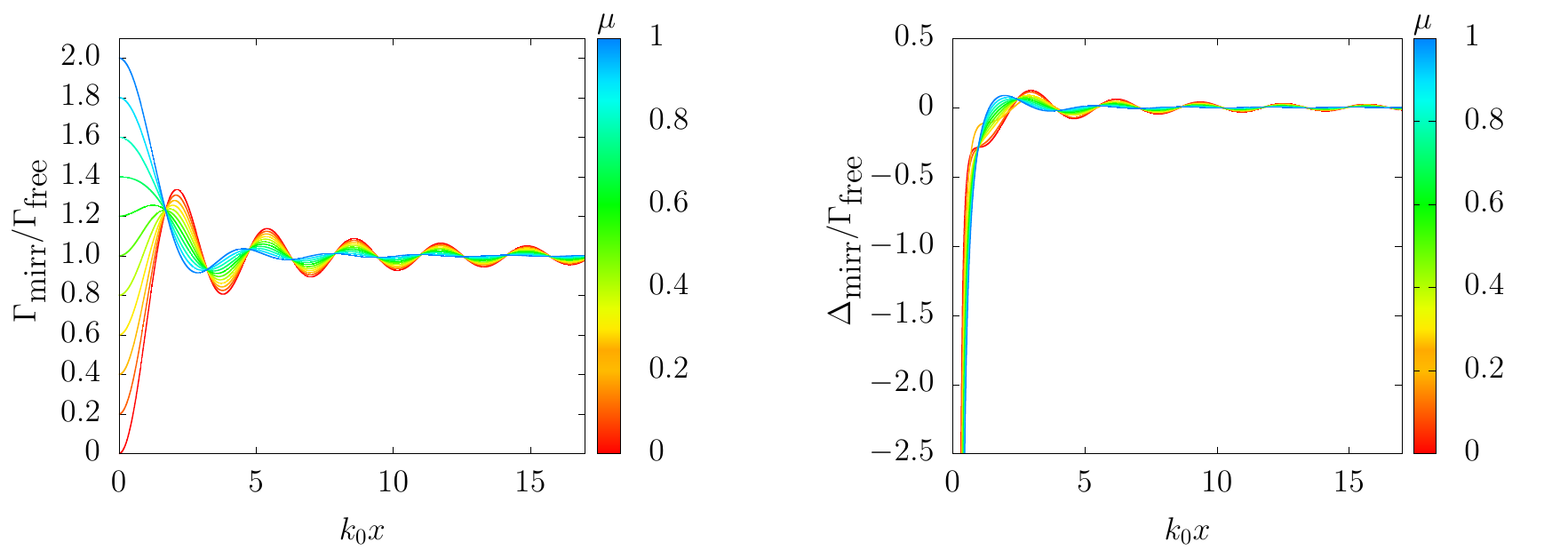} 
    \caption{[Colour online] The spontaneous decay rate $\Gamma_{\rm mirr}$ (a) and the atomic level shift $\Delta_{\rm mirr}$ (b) of an atom in front of a perfect mirror (c.f.~Eq.~\eqref{eq:rho12d}) as a function of the atom-mirror distance $x$ for different  orientations of the atomic dipole moment ${\bf D}_{12}$. For $\mu=0$, ${\bf D}_{12}$ is parallel and, for $\mu = 1$, ${\bf D}_{12}$ is perpendicular to the mirror surface. In all cases, we have $\Gamma_{\rm mirr} = 0$ while $\Delta_{\rm mirr}$ diverges for $x=0$. Moreover, for $k_{0} x \gg 1$, we have $\Gamma_{\rm mirr} = \Gamma_{\rm free}$ and $\Delta_{\rm mirr} = 0 $, as it should. For plot (a), the $\mu=0$ case corresponds to the line emanating from $0$ and becomes the maximum of the oscillation, where as the $\mu=1$ case corresponds to the line emanating from $\Gamma_{\rm mirr} = 2 \Gamma_{\rm free}$ is increased and becomes the minimum of the oscillation. For plot (b), the $\mu=0$ case corresponds to the line on left-hand side of the graph (line emanating from minus infinity) which gives the most noticeable shift in the atom's energy levels, where as the $\mu=1$ case corresponds to the line on right-hand side which gives the least noticeable level shift.}
    \label{fig:AM1}
\end{figure*}

In the following we use the Hamiltonian $H_{\rm I}(t)$ in Eq.~\eqref{eq:E(x)45} and standard quantum optical  approximations to evaluate $H_{\rm cond}$ and ${\cal L} (\rho_{\rm AI} (t))$ in Eqs.~(\ref{eq:rho4}) and (\ref{eq:rho4xxx}). Doing so and proceeding as described in Apps.~\ref{appC} and \ref{appD}, we find that
\begin{eqnarray} \label{eq:rho9}
H_{\rm cond} &=& \hbar \left( \Delta_{\rm mirr} - \frac{{\rm i}}{2} \, \Gamma_{\rm mirr} \right) \sigma^+ \sigma^- \, , \nonumber \\
{\cal L} (\rho_{\rm AI} (t)) &=& \Gamma_{\rm mirr} \, \sigma^- \, \rho_{\rm AI} (t) \, \sigma^+ 
\end{eqnarray}
with
\begin{widetext}
\begin{eqnarray} \label{eq:rho12}
\Gamma_{\rm mirr} &=& \left[ {1 + r_{a}^{ 2} \over \eta_{a}^{2}} + {t_{b}^{ 2} \over \eta_{b}^{2}} \right] \Gamma_{\rm free}  - {3r_{a} \over \eta_{a}^{2}} \left[ {\sin (2 k_0x) \over 2k_{0}x} \left( 1 - \mu \right) + \left( {\cos (2 k_0x) \over (2k_0x)^2} - {\sin (2 k_0x) \over (2k_0x)^3} \right) \left( 1 + \mu \right) \right] \Gamma_{\rm free} \, , \nonumber \\
\Delta_{\rm mirr} &=& {3 r_{a} \over 2 \eta_{a}^{2}} \left[ {\cos (2 k_{0} x) \over 2 k_{0} x} (1-\mu) - \left( {\sin (2 k_{0} x) \over (2 k_{0} x)^{2}} + {\cos (2 k_{0} x) \over (2 k_{0} x)^{3}} \right) (1+\mu) \right] \Gamma_{\rm free} \, . 
\end{eqnarray}
\end{widetext}
Here $k_{0} = \omega_{0}/c$ and the constant $\mu$ denotes the orientation of the atomic dipole moment,
\begin{eqnarray} \label{eq:rho12b}
\mu &=& \| \hat {\bf D}_{12} \cdot \hat {\bf x} \|^2 
\end{eqnarray}
with $\hat {\bf x}$ and $\hat {\bf D}_{12}$ being unit vectors. The above equations show that the presence of the mirror alters the spontaneous decay rate $\Gamma_{\rm mirr}$ of an atom near a semi-transparent mirror and causes a level shift $\Delta_{\rm mirr}$ of the excited atomic state $|2 \rangle$. When deriving the above level shift, we neglect a self-interaction term which is also present in free space. This term is independent of the presence of the mirror and can be absorbed into the definition of $\omega_0$ \cite{Stokes}.

\begin{figure*}[t]
    \centering
		\includegraphics[width=0.98\textwidth]{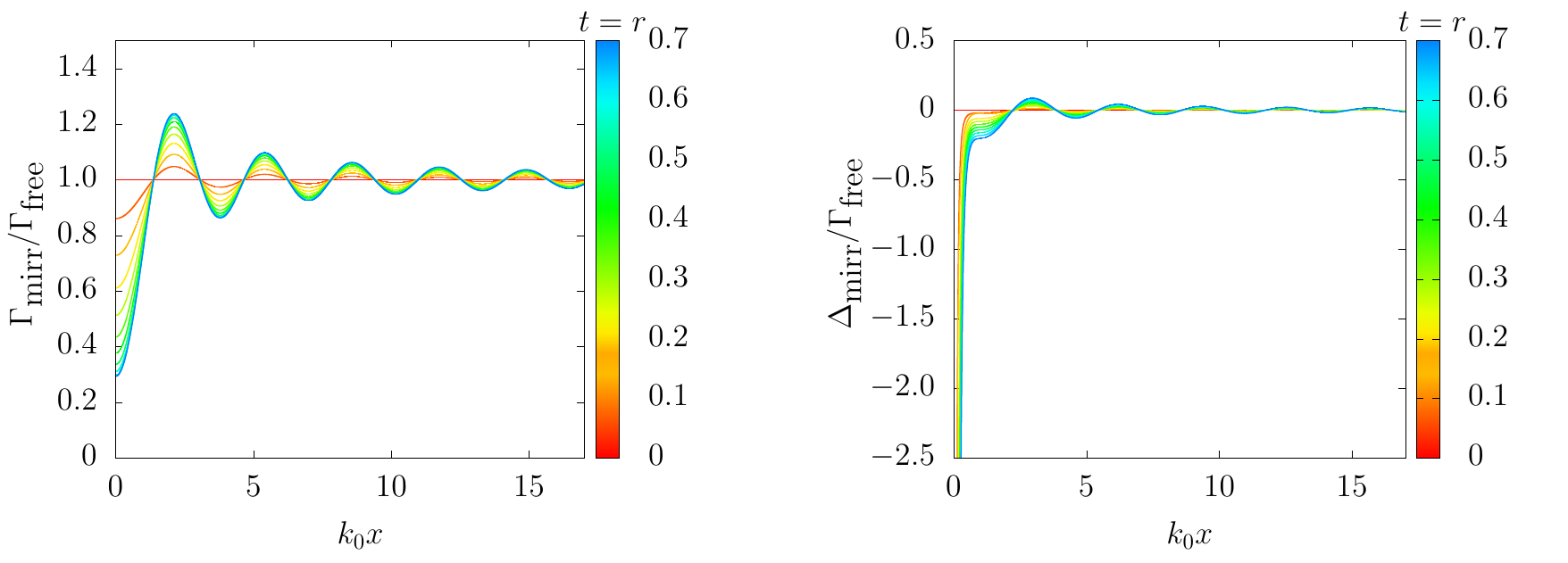} 
\caption{[Colour online] The spontaneous decay rate $\Gamma_{\rm mirr}$ (a) and the atomic level shift $\Delta \omega_{\rm mirr}$ (b) of an atom in front of a symmetric mirror (c.f.~Eq.~\eqref{eq:rho12zz}) as a function of the atom-mirror distance $x$ for different values of $r$. For simplicity we assume here that $r = t$ and $\mu = 0$. The case $r=0$ corresponds to a completely absorbing surface, while $r = 1/\sqrt{2}$ models a $50:50$ beamsplitter.}
	\label{fig:AM2}
\end{figure*}

For atom-mirror distances $x$ much larger than the wavelength $\lambda_0$ of the emitted light, we have $k_0 x \gg 1$ and Eqs.~\eqref{eq:rho12} simplify to $\Delta_{\rm mirr} = 0$, while
\begin{eqnarray} \label{eq:rho12c}
\Gamma_{\rm mirr} &=& \left[ {1 + r_{a}^{ 2} \over \eta_{a}^{2}} + {t_{b}^{ 2} \over \eta_{b}^{2}} \right] \Gamma_{\rm free} \, .
\end{eqnarray}
Analogously one can show that the spontaneous decay rate of an atom on the left side of the mirror equals 
\begin{eqnarray} \label{eq:rho12z}
\Gamma_{\rm mirr} &=& \left[ {1 + r_{b}^{ 2} \over \eta_{b}^{2}} + {t_{a}^{ 2} \over \eta_{a}^{2}} \right] \Gamma_{\rm free} 
\end{eqnarray}
when $k_0 |x| \gg 1$. Assuming that the mirror borders on both sides on a medium with permittivity $\varepsilon$ and demanding that these decay rates both coincide with the free-space decay rate $\Gamma_{\rm free}$ finally yields the normalisation factors $\eta_a$ and $\eta_b$. In free space, where we have zero reflection $(r_{a} = r_{b} = 0)$ and maximum transmission $(t_{a} = t_{b}=1)$, this applies when
\begin{eqnarray} \label{eq:final3}
{1 \over \eta_a^2} + {1 \over \eta_b^2} = 1 \, . 
\end{eqnarray}
However, in general, $\eta_a$ and $\eta_b$ are given by
\begin{eqnarray} \label{final10}
\eta_a^2 &=& {\left( 1 + r_a^{ 2} \right) \left(1 + r_b^{ 2} \right) - (t_a t_b)^2 \over 1 + r_b^{ 2} - t_b^{ 2}} \, , \nonumber \\
\eta_b^2 &=& {\left( 1 + r_a^{ 2} \right) \left(1 + r_b^{ 2} \right) - (t_a t_b)^2 \over 1 + r_a^{ 2} - t_a^{ 2}} \, .
\end{eqnarray}
The (real) transmission and reflection rates in these equations can be determined experimentally.

\subsubsection{Perfect mirrors}

To gain more intuition for the results in Eqs.~(\ref{eq:rho12}), we now have a closer look at some concrete scenarios. For example, in the case of a perfect mirror, we have $r_{a} = r_{b} = 1$ and $t_{a} = t_{b}=0$. Using Eq.~(\ref{final10}), we find that this implies $\eta_a = \eta_b = \sqrt{2}$, as stated in Eq.~(\ref{final6}). Substituting these parameters into Eq.~\eqref{eq:rho12} yields
\begin{eqnarray} \label{eq:rho12d}
\Gamma_{\rm mirr} &=& \Gamma_{\rm free} - {3  \over 2} \left[ {\sin (2 k_0x) \over 2k_0x} \left( 1 - \mu \right) \right. \nonumber \\
&& \left. + \left( {\cos (2 k_0x) \over (2k_0x)^2} - {\sin (2 k_0x) \over (2k_0x)^3} \right) \left( 1 + \mu \right) \right] \Gamma_{\rm free} \, , \nonumber \\
\Delta_{\rm mirr} &=& {3 \over 4} \left[ {\cos (2 k_{0} x) \over (2 k_{0} x)} (1-\mu) \right. \nonumber\\
&& \hspace*{0cm} \left. - \left( {\sin (2 k_{0} x) \over (2 k_{0} x)^{2}} + {\cos (2 k_{0} x) \over (2 k_{0} x)^{3}} \right) (1+\mu) \right] \Gamma_{\rm free}  \nonumber\\
\end{eqnarray}
which are in good agreement with the findings of other authors (see e.g.~Ref.~\cite{Morawitz}). The altered spontaneous decay rate $\Gamma_{\rm mirr}$ and the level shift $\Delta_{\rm mirr}$ are the result of a dipole-dipole interaction between the atom and its mirror image \cite{dipdip}. However, as one can see from Eq.~\eqref{eq:E(x)45}, the $x$-component of the atomic dipole moment of the mirror image, ${\widetilde {\bf D}}_{12}$, and the dipole moment ${\bf D}_{12}$ of the original atom have different signs.

Fig.~\ref{fig:AM1} shows the $x$ dependence of the spontaneous decay rate $\Gamma_{\rm mirr}$ and the level shift $\Delta_{\rm mirr}$ of an atom in front of a perfect mirror for different dipole orientations $\mu$. For distances $x$ of the same order of magnitude as the wavelength $\lambda_0$ of the emitted light, the last terms in Eq.~\eqref{eq:rho12d} are no longer negligible and $\Gamma_{\rm mirr}$ and $\Delta_{\rm mirr}$ both depend strongly on $x$ and $\mu$. As one would expect, this dependence is most pronounced and most long-range when $\mu = 0$, i.e.~in the case of an atomic dipole moment that is parallel to the mirror surface. In contrast to this, the decay rate $\Gamma_{\rm mirr}$ approaches $\Gamma_{\rm free}$ much more quickly when $\mu = 1$. In both cases, we have $\Gamma_{\rm mirr} = 0$ for $x=0$, since the electric field amplitude vanishes on the surface of a perfectly conducting mirror (c.f.~Eq.~\eqref{eq:boundarycondition}).

\subsubsection{Symmetric mirrors}

\begin{figure*}[t]
    \centering
		\includegraphics[width=0.98\textwidth]{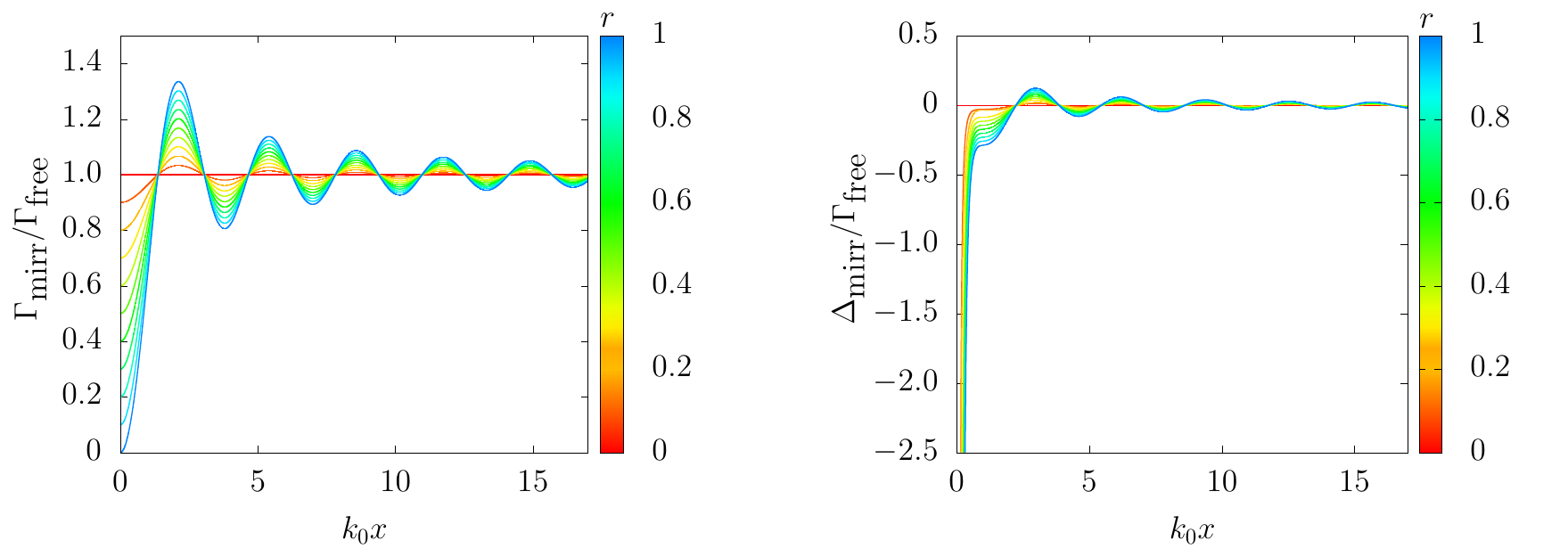} 
\caption{[Colour online] The spontaneous decay rate $\Gamma_{\rm mirr}$ (a) and the atomic level shift $\Delta_{\rm mirr}$ (a) of an atom in front of a non-absorbing symmetric mirror (c.f.~Eqs.~\eqref{eq:rho12zz}) as a function of the atom-mirror distance $x$ for different values of $r$. Again we assume $\mu=0$, while $t^2 = 1- r^2$. The case $r=0$ corresponds to free space, while $r = 1$ models a perfect mirror. For plot (a), the $r=0$ case corresponds to the horizontal line emanating from $\Gamma_{\rm mirr}/\Gamma_{\rm free}=1$ meaning that the atom decays as in free space when there is no mirror present, where as the $r=1$ case corresponds to the line emanating from $0$ which gives the maximum of the oscillation. For plot (b), the $r=0$ case corresponds to the horizontal line emanating from $\Delta_{\rm mirr}/\Gamma_{\rm free}=0$ meaning that there is no mirror-induced shift in the atomic energy levels. The $r=1$ case corresponds to the most noticeable shift in the atomic energy levels.}
\label{fig:AM3}
\end{figure*}

In the case of a symmetric mirror with equal transmission and reflection rates, i.e.~when 
\begin{eqnarray}
t_a = t_b = t ~~ &{\rm and}&~~ r_a = r_b = r \, , 
\end{eqnarray}
the normalisation factors $\eta_a$ and $\eta_b$ in Eq.~\eqref{final10} become the same. In this case we have $\eta_a^2 = \eta_b^2 = 1+2r^2$. Hence, using Eq.~\eqref{eq:rho12}, one can show that 
\begin{eqnarray} \label{eq:rho12zz}
\Gamma_{\rm mirr} &=& \Gamma_{\rm free} - { 3 r \left(1+r^2-t^2 \right) \over (1+r^{2})^{2} - t^{4}} \, \left[ {\sin (2 k_0x) \over 2k_0x} \left( 1 - \mu \right) \right. \nonumber\\
&& \left. + \left( {\cos (2 k_0x) \over (2k_0x)^2} - {\sin (2 k_0x) \over (2k_0x)^3} \right) \left( 1 + \mu \right) \right] \, \Gamma_{\rm free} \, ,  ~~~ \nonumber \\
\Delta_{\rm mirr} &=&{ 3 r \left(1+r^2-t^2 \right) \over (1+r^{2})^{2} - t^{4}} \bigg[ {\cos (2 k_0x) \over 2k_0x} \left( 1 - \mu \right) \bigg. \nonumber\\
&& \bigg. - \left( {\sin (2 k_0x) \over (2k_0x)^2} + {\cos (2 k_0x) \over (2k_0x)^3} \right) \left( 1 + \mu \right) \bigg] \, \Gamma_{\rm free} \nonumber \\
\end{eqnarray}
in this case. Again, $\Gamma_{\rm mirr}$ and $\Delta_{\rm mirr}$ depend strongly on $\mu$ and $r$ for relatively short atom-mirror distances $x$ but tend to their respective free-space rates when $x$ becomes much larger than $\lambda_0$. This is illustrated in Figs.~\ref{fig:AM2} and \ref{fig:AM3} which show $\Gamma_{\rm mirr}$ and and $\Delta_{\rm mirr}$ as a function of $x$ for different values of $r$ and $t$, while $\mu = 0$. 

\subsubsection{Highly-absorbing mirrors}

Finally, we have a closer look at a mirror that absorbs all incoming light. This case is equivalent to a perfectly-transmitting mirror with $r_a = 0$ which yields (c.f.~Eq.~\eqref{eq:rho12})
\begin{eqnarray} \label{eq:rho12zzz}
\Gamma_{\rm mirr} = \Gamma_{\rm free} ~~ & {\rm and} & ~~ \Delta_{\rm mirr} = 0 
\end{eqnarray}
independent of the atom-mirror distance $x$ and the orientation $\mu$ of the atomic dipole moment. As one would expect, an atom near an absorbing medium does not see the mirror and decays exactly as it would in free space. This is illustrated in Figs.~\ref{fig:AM2} and \ref{fig:AM3} which both show a flat line for $r_a = 0$.

\section{Conclusions} \label{conclusions}

The main result of this paper is the quantisation of the electromagnetic field near a semi-transparent mirror. Using an image detector method, we obtain expressions for the system Hamiltonian $H_{\rm sys}$ of field and mirror and the electric field observable ${\bf E}_{\rm mirr} ({\bf r})$ (c.f.~Eqs.~\eqref{eq:Hfree2222} and \eqref{eq:38full} with $\eta_a$ and $\eta_b$ as in Eq.~\eqref{final10}). In contrast to $H_{\rm sys}$, which is independent of the transmission and reflection rates $t_a$, $t_b$, $r_a$ and $r_b$ of the mirror, the electric field observable ${\bf E}_{\rm mirr} ({\bf r})$ depends strongly on these rates. The possible absorption of light in the mirror surface is explicitly taken into account, since the squares of the absorption and transmission coefficients, i.e.~$r_a^{ 2} + t_a^{ 2}$ and $r_b^{ 2} + t_b^{ 2}$, do not have to add up to one. However, for simplicity we assume in this paper that the reflection and the transmission rates of the mirror do not depend on the frequency and the angle of the incoming light.

Before quantising the electromagnetic field, Sec.~\ref{CIM} uses classical electrodynamics to discuss the scattering of light on flat surfaces. One way of modelling the scattering process is to assume that incoming wave packets evolve exactly as in free space. However, the presence of the mirror changes how and where the amplitudes of the electromagnetic field are measured. Adopting this point of view when deriving the observables of the electromagnetic field near a two-sided semi-transparent mirror, we find that the system Hamiltonian $H_{\rm sys}$ is the sum of two free-space field Hamiltonians $H_{\rm free}$. Moreover, ${\bf E}_{\rm mirr} ({\bf r})$ is now a sum of electric field free-space observables which can be associated with incoming, reflected and transmitted waves. To correctly normalise these contributions, we demand that an atom at a relatively large distance away from the mirror has the same spontaneous decay rate as an atom in free space. In addition, phase factors need to be introduced such that maximum interference on one side of the mirror implies minimum interference on the other. Our field observables have some similarities with previously proposed observables \cite{Carniglia,Zakowicz,Wu,Eberlein09,Eberlein12,Hammer} but can be used to model not only one-sided but two-sided semi-transparent mirrors. 

Another difference between our field quantisation scheme and the schemes of other authors is that the energy of the mirror surface, i.e.~the energy of the mirror images, is explicitly taken into account. For example, when placing a single wave packet in front of a one-sided perfect mirror, half of the energy of the system belongs to the original wave packet and the other half belongs to its mirror image  and is stored in mirror surface charges. In general, there is a difference between the system Hamiltonian $H_{\rm sys}$ and the Hamiltonian $H_{\rm field}$ of the electromagnetic field surrounding the semi-transparent mirror. Energy can flow from the field onto the mirror surface and back. In the case of absorption, the interaction with the mirror surface reduces the energy of incoming wave packets without changing their shape.

Finally, to test the consistency of our model in limiting cases and to determine some previously unknown normalisation factors, we derive the spontaneous decay rate $\Gamma_{\rm mirr}$ and the level shift $\Delta_{\rm mirr}$ of an atom in front of a two-sided semi-transparent mirror. In good agreement with what one would expect, highly absorbing mirrors do not alter the spontaneous decay rate of the atom. However, in general, $\Gamma_{\rm mirr}$ and $\Delta_{\rm mirr}$ depend in a relatively complex way on transmission and reflection rates and other relevant system parameters (c.f.~Eq.~\eqref{eq:rho12}). We expect that the results derived in this paper have a wide range of applications, for example, when designing novel photonic devices for quantum technology applications. \\[0.5cm]
{\em Acknowledgement.} We would especially like to thank Robert Bennett, Axel Kuhn, Thomas Mann and Jake Southall for fruitful and stimulating discussions. Moreover, we acknowledge financial support from the Oxford Quantum Technology Hub NQIT (grant number EP/M013243/1) and the EPSRC (award number 1367108). Statement of compliance with EPSRC policy framework on research data: This publication is theoretical work that does not require supporting research data.

\appendix
\section{Derivation of the constraint in Eq.~(\ref{phases})} \label{appphases}

Suppose two relatively well-localised wave packets approach a two-sided semi-transparent mirror from either side. In the following we consider only one specific frequency contribution of these wave packets with positive wave number $k$ and with
\begin{eqnarray}
E_{\rm mirr}^{(a)}(x,0) &=& \left[ E_0^{(a)} \, {\rm e}^{{\rm i} \xi_1} \, {\rm e}^{- {\rm i} kx} + {\rm c.c.} \right] \, \Theta(x) \, , \nonumber \\
E_{\rm mirr}^{(b)}(x,0) &=& \left[ E_0^{(b)} \, {\rm e}^{{\rm i} \xi_2}  \, {\rm e}^{{\rm i} kx} + {\rm c.c.} \right] \, \Theta(-x) \, ,
\end{eqnarray}
where $E_0^{(a)}$ and $E_0^{(b)}$ denote real amplitudes and $\xi_1$ and $\xi_2$ specify initial phases. After a sufficiently long time, once both wave packets have seen the mirror, the electric field $E_{\rm mirr}(x,t)$ is given by
\begin{eqnarray}
E_{\rm mirr}(x,t) &=& \left[ r_a \, E_0^{(a)} \, {\rm e}^{{\rm i} (\xi_1 + \varphi_1)} \, {\rm e}^{{\rm i} (kx - \omega t)} \right. \nonumber \\
&& \left. + t_b \, E_0^{(b)} \, {\rm e}^{{\rm i} (\xi_2 + \varphi_2)} \, {\rm e}^{{\rm i} (kx - \omega t)} \right] \, \Theta(x) \, , \nonumber \\
&& + \left[ r_b \, E_0^{(b)} \, {\rm e}^{{\rm i} (\xi_2 + \varphi_3)} \, {\rm e}^{- {\rm i} (kx + \omega t)} \right. \nonumber \\
&& \left. + t_a \, E_0^{(a)} \, {\rm e}^{{\rm i} (\xi_1 + \varphi_4)} \, {\rm e}^{- {\rm i} (kx + \omega t)} \right] \, \Theta(-x) \nonumber \\
&& + {\rm c.c.} \, ,
\end{eqnarray}
which is in agreement with Eq.~(\ref{eq:E(x)24}). Rearranging this equation, we find that $E_{\rm mirr}(x,t)$ also equals
\begin{eqnarray}
E_{\rm mirr}(x,t) &=& \left[ r_a \, E_0^{(a)} + t_b \, E_0^{(b)} \, {\rm e}^{{\rm i} (\xi_2 - \xi_1 + \varphi_2 - \varphi_1)}  \right] \nonumber \\
&& \times {\rm e}^{{\rm i} (\xi_1 + \varphi_1)} \, {\rm e}^{{\rm i} (kx - \omega t)} \, \Theta(x) \, , \nonumber \\
&& + \left[ t_a \, E_0^{(a)}  + r_b \, E_0^{(b)} \, {\rm e}^{{\rm i} (\xi_2 - \xi_1 + \varphi_3 - \varphi_4)} \, \right] ~~~ \nonumber \\ 
&& \times {\rm e}^{{\rm i} (\xi_1 + \varphi_4)} \, {\rm e}^{- {\rm i} (kx + \omega t)} \, \Theta(-x) + {\rm c.c.} 
\end{eqnarray}
which shows that maximum interference of electric field amplitudes on one side of the mirror {\em always}  implies minimum interference on the other side, when
\begin{eqnarray}
{\rm e}^{{\rm i} (\xi_2 - \xi_1 + \varphi_2 - \varphi_1)}  &=& - {\rm e}^{{\rm i} (\xi_2 - \xi_1 + \varphi_3 - \varphi_4)} \, .
\end{eqnarray}
This equation yields Eq.~(\ref{phases}) in the main text. The same applies for the magnetic field amplitudes which interfere in the same way on the same side of the mirror, as the electric field amplitudes.

\section{Calculation of $H_{\rm field}$ for a one-sided perfect mirror} \label{AppendixA2}

Substituting the electromagnetic field observables $E_{\rm mirr}(x)$ and $B_{\rm mirr}(x)$ in Eq.~\eqref{eq:E(x)32b} into Eq.~\eqref{eq:Hfree12}, we find that 
\begin{eqnarray} \label{eq:HfieldA1later}
H_{\rm field} &=& - \frac{\hbar}{8 \pi}  \, \int_0^{\infty} {\rm d} x \, \int_{-\infty}^{\infty} {\rm d} k \, \int_{-\infty}^{\infty} {\rm d} k^{\prime} \, \sqrt{\omega \omega^{\prime}} \nonumber\\
&& \times \left( {\rm e}^{{\rm i}kx} \xi_{k} - {\rm e}^{-{\rm i}kx} \xi_{k}^{\dagger} \right)
\left( {\rm e}^{{\rm i}k^{\prime}x} \xi_{k^{\prime}} - {\rm e}^{- {\rm i}k^{\prime}x} \xi_{k^{\prime}}^{\dagger} \right) \nonumber\\
&& \times \left[ 1 + \text{sign}(kk') \right] \, . ~~~
\end{eqnarray}
Replacing $x$ by $-x$, $k$ by $-k$ and $k'$ by $-k'$, and taking into account that $\xi_{-k} = - \xi_k$ by definition, one can show that this field Hamiltonian can also be written as 
\begin{eqnarray} \label{eq:HfieldA1later2}
H_{\rm field} &=& - \frac{\hbar}{8 \pi}  \, \int_{-\infty}^0 {\rm d} x \, \int_{-\infty}^{\infty} {\rm d} k \, \int_{-\infty}^{\infty} {\rm d} k^{\prime} \, \sqrt{\omega \omega^{\prime}} \nonumber\\
&& \times \left( {\rm e}^{{\rm i}kx} \xi_{k} - {\rm e}^{-{\rm i}kx} \xi_{k}^{\dagger} \right)
\left( {\rm e}^{{\rm i}k^{\prime}x} \xi_{k^{\prime}} - {\rm e}^{- {\rm i}k^{\prime}x} \xi_{k^{\prime}}^{\dagger} \right) \nonumber\\
&& \times \left[ 1 + \text{sign}(kk') \right] \, . ~~~
\end{eqnarray}
Adding both operators, we find that
\begin{eqnarray} \label{eq:HfieldA1later}
H_{\rm field} &=& - \frac{\hbar}{16 \pi}  \, \int_{-\infty}^{\infty} {\rm d} x \, \int_{-\infty}^{\infty} {\rm d} k \, \int_{-\infty}^{\infty} {\rm d} k^{\prime} \, \sqrt{\omega \omega^{\prime}} \nonumber\\
&& \times \left( {\rm e}^{{\rm i}kx} \xi_{k} - {\rm e}^{-{\rm i}kx} \xi_{k}^{\dagger} \right)
\left( {\rm e}^{{\rm i}k^{\prime}x} \xi_{k^{\prime}} - {\rm e}^{- {\rm i}k^{\prime}x} \xi_{k^{\prime}}^{\dagger} \right) \nonumber\\
&& \times \left[ 1 + \text{sign}(kk') \right] \, . ~~~
\end{eqnarray}
Finally, we employ the relation
\begin{eqnarray} \label{eq:Diracdelta}
\int_{-\infty}^{\infty} {\rm d} x \, {\rm e}^{\pm{\rm i} k_0 x} &=& 2 \pi \, \delta(k_0) \, , ~~
\end{eqnarray}
where $k_0$ denotes a constant, to show that
\begin{eqnarray} \label{eq:HfieldA4later}
H_{\rm field} &=& \int_{- \infty}^{\infty} {\rm d} k \, \frac{1}{4} \hbar \omega \, \left( \xi_{k} \xi_{k}^{\dagger} + \xi_{k}^{\dagger} \xi_{k} \right) \, . ~~
\end{eqnarray}
This Hamiltonian differs from the field Hamiltonian in Eq.~\eqref{final} only by a constant summand.

\section{Calculation of $H_{\rm cond}$ for a semi-transparent mirror} \label{appC}

Combining Eqs.~\eqref{eq:rho4} and \eqref{eq:E(x)45}, we find that the conditional Hamiltonian $H_{\rm cond}$ of an atom in front of a semi-transparent mirror equals
\begin{eqnarray} \label{eq:rho6}
H_{\rm cond} &=& - \int \limits_{t}^{t + \Delta t} {\rm d}t^{\prime} \int \limits_{t}^{t^{\prime}} {\rm d}t^{\prime \prime} \sum_{\lambda = 1,2}  \int_{\mathbb{R}^{3}} {\rm d}^3 {\bf k} \, \frac{{\rm i} e^2 \omega}{16 \pi^3 \varepsilon \, \Delta t} \nonumber \\
&& \times \left[ {1 \over \eta_a^2} \left\| \left( {\bf D}^{\star}_{12} \,  {\rm e}^{{\rm i} {\bf k} \cdot {\bf r}} - r_a \, \widetilde{\bf D}^{\star}_{12} \, {\rm e}^{{\rm i} {\bf k} \cdot \widetilde{\bf r}} \right) \cdot \hat {\bf e}_{{\bf k} \lambda} \right\|^2 \right. \nonumber \\
&& + \left. {t_b^{ 2} \over \eta_b^2} \left\| \left( \mathbf{D}^{\star}_{12} \cdot \hat {\bf e}_{{\bf k} \lambda} \right) \right\|^2 \right] {\rm e}^{- {\rm i} (\omega - \omega_0) (t' - t'')} \, \sigma^+ \sigma^- . \nonumber \\
\end{eqnarray}
Without restrictions, we consider in the following a co-ordinate system in which the atomic dipole moment ${\bf D}_{12}$ can be written as
\begin{eqnarray}
{\bf D}_{12} &=&  \| {\bf D}_{12} \| \, \left( \begin{array}{c} d_1 \\ 0 \\ d_3 \end{array} \right)
\end{eqnarray}
with $|d_1|^2 + |d_3|^2 = 1$. In this coordinate system, the dipole moment $\widetilde {\bf D}_{12}$ of the mirror image of the atom equals
\begin{eqnarray} \label{Alice}
\widetilde{\bf D}_{12} &=&  \| {\bf D}_{12} \| \, \left( \begin{array}{c} - d_1 \\ ~ 0 \\ ~ d_3 \end{array} \right) \, .
\end{eqnarray}
To simplify Eq.~\eqref{eq:rho6}, we notice that the polarisation vectors $\hat {\bf e}_{{\bf k} \lambda}$ with $\lambda = 1,2$ and the unit vector $\hat {\bf k} = {\bf k}/\|\mathbf{k}\|$ form a complete set of basis states in $\mathbb{R}^{3}$ which implies
\begin{eqnarray} \label{eq:rho88}
\sum_{\lambda = 1,2} \left\| {\bf v} \cdot \hat {\bf e}_{{\bf k} \lambda} \right\|^2
&=& \| {\bf v} \|^2 - \| {\bf v} \cdot \hat{\bf k} \|^2  
\end{eqnarray}
for any vector ${\bf v}$. Moreover, to perform the integration in ${\bf k}$-space, we introduce the polar coordinates $(\omega, \varphi, \vartheta)$ such that 
\begin{eqnarray}
{\bf k} &=& {\omega  \over c} \, \left( \begin{array}{c} \cos (\vartheta) \\ \cos (\varphi) \, \sin (\vartheta) \\ \sin (\varphi) \, \sin (\vartheta) \end{array} \right)
\end{eqnarray}
and 
\begin{eqnarray} \label{eq:rho888}
\int_{\mathbb{R}^{3}} {\rm d}^3 {\bf k} &=& \int_{0}^\infty {\rm d} \omega \int_0^{\pi} {\rm d} \vartheta \int_0^{2\pi} {\rm d} \varphi \, {\omega^2 \over c^3} \, \sin (\vartheta) \, . ~~
\end{eqnarray}
Defining $\mu$ as in Eq.~(\ref{eq:rho12b}) implies $|d_1|^2 = \mu$ and $|d_3|^2 = 1- \mu$. Taking this into account and 
combining the above equations, performing the $\varphi $ integration and introducing two new variables, $s = \cos (\vartheta) $ and $\xi = t' - t''$, hence yields
\begin{eqnarray} \label{eq:rho62}
H_{\rm cond} &=& - \int \limits_{t}^{t + \Delta t} {\rm d}t^{\prime} \int \limits_0^{t^{\prime}- t} {\rm d}\xi \int_{0}^\infty {\rm d} \omega \int_{-1}^1 {\rm d} s \, {{\rm i} e^2 \omega^3 \left\| {\bf D}_{12} \right\|^2 \over 8 \pi^2  \varepsilon c^3 \, \Delta t} \nonumber \\
&& \times \left[ {1 \over \eta_a^2} \left(1 + r_a^{ 2} + 2 r_a \cos \left( 2kxs \right) \right) \left(1 - s^2 \right) \mu \right. \nonumber \\
&& + {1 \over 2 \eta_a^2} \left(1 + r_a^{ 2} - 2 r_a \cos \left( 2kxs \right) \right) \left(1 + s^2 \right) (1-\mu) \nonumber \\
&& \left. + {t_b^{ 2} \over \eta_b^2} \left(1 - s^2 \right) \mu + {t_b^{ 2} \over 2 \eta_b^2} \left(1 + s^2 \right) (1-\mu) \right] \nonumber \\
&& \times {\rm e}^{- {\rm i} (\omega - \omega_0) \xi} \, \sigma^+ \sigma^- 
\end{eqnarray}
with $k=\omega/c$. Next we extend the $\xi$-integral to infinity. This is well justified when $t^{\prime} - t$ is similar to $\Delta t$ and $\Delta t \gg 1/\omega_0$, as is in general the case. Doing so and performing the $t'$ and the $s$ integration, we obtain the conditional Hamiltonian
\begin{eqnarray} \label{eq:rho9later}
H_{\rm cond} &=& \hbar C_{\rm mirr} \, \sigma^+ \sigma^- 
\end{eqnarray}
with the constant $C_{\rm mirr}$ given by 
\begin{eqnarray} \label{eq:C10}
C_{\rm mirr} &=& - {{\rm i} \over 2 \pi} \, {\Gamma_{\rm free} \over \omega_0^3} \int \limits_0^{\infty} {\rm d}\xi \int_{0}^\infty {\rm d} \omega \, \omega^3 {\rm e}^{- {\rm i} (\omega - \omega_0) \xi} \nonumber \\
&& \hspace*{-0.2cm} \times \left[ {1 + r_a^{ 2} \over \eta_a^2} + {t_b^{ 2} \over \eta_b^2} - {3 r_a \over \eta_a^2} \, {\sin \left( 2kx \right) \over 2kx} (1-\mu) \right. \nonumber \\
&& \hspace*{-0.2cm} \left. - {3 r_a \over \eta_a^2} \left( {\cos \left( 2kx \right) \over (2kx)^2} - {\sin \left( 2kx \right) \over (2kx)^3} \right) (1 + \mu) \right] ~~~
\end{eqnarray}
with the spontaneous free space decay rate $\Gamma_{\rm free}$ defined as in Eq.~(\ref{Gamma0}). Next we perform the $\xi$ integration by taking into account that 
\begin{eqnarray} \label{eq:rho64}
\int \limits_0^{\infty} {\rm d}\xi \, {\rm e}^{- {\rm i} (\omega - \omega_0) \xi} 
&=& \pi \, \delta (\omega - \omega_0) + {{\rm i} \over \omega - \omega_0} \, . ~~
\end{eqnarray}
Doing so, one can show that
\begin{widetext}
\begin{eqnarray} \label{eq:C11}
C_{\rm mirr} &=& - {{\rm i} \over 2} \left[ {1 + r_a^{ 2} \over \eta_a^2} + {t_b^{ 2} \over \eta_b^2} \right] \, \Gamma_{\rm free} 
- {{\rm i} \over 2} \left( - {3 r_a \over \eta_a^2} \right) \left[  {\sin \left( 2k_0x \right) \over 2k_0x} (1-\mu) + \left( {\cos \left( 2k_0x \right) \over (2k_0x)^2} - {\sin \left( 2k_0x \right) \over (2k_0x)^3} \right) (1 + \mu) \right] \, \Gamma_{\rm free} \nonumber \\
&& - {1 \over 2\pi} \, {3r_a \over \eta_a^2} \int_{0}^\infty {\rm d} \omega \, {\omega^3 \over \omega - \omega_0} \, \left[ {\sin \left( 2kx \right) \over 2kx} (1-\mu) - \left( {\cos \left( 2kx \right) \over (2kx)^2} - {\sin \left( 2kx \right) \over (2kx)^3} \right) (1+ \mu) \right] \, {\Gamma_{\rm free} \over \omega_0^3} \nonumber\\
&& + {1 \over 2 \pi} \, \left[  {1 + r_a^{ 2} \over \eta_a^2} + {t_b^{ 2} \over \eta_b^2} \right] \int_{0}^\infty {\rm d} \omega \, {\omega^3 \over \omega - \omega_0} \, {\Gamma_{\rm free} \over \omega_0^3} \, .
\end{eqnarray}
\end{widetext}
From the general form of the conditional Hamiltonian in Eq.~(\ref{eq:rho9}) we see that the imaginary part of this constant denotes a spontaneous decay rate, while its real part denotes an atomic level shift \cite{Stokes,He1}. More concretely, comparing Eqs.~(\ref{eq:rho9}) and (\ref{eq:rho9later}), we find that 
\begin{eqnarray}
&& \hspace*{-0.5cm} \Gamma_{\rm mirr} = - 2 \, {\rm Im} \, C_{\rm mirr} \, , ~~ \Delta_{\rm mirr} = {\rm Re} \, C_{\rm mirr} \, . ~~
\end{eqnarray}
Demanding that $\Gamma_{\rm mirr}$ equals $\Gamma_{\rm free}$ for large values of $x$ shows that the square bracket in the last line of Eq.~(\ref{eq:C11}) equals unity (c.f.~Eq.~(\ref{final10})). The last term therefore describes an atomic level shift which does not depend on the presence of the mirror. As usual, we absorb this level shift in the following into the definition of $\omega_0$, thereby absorbing it into the atomic Hamiltonian $H_{\rm atom}$ in Eq.~(\ref{eq:Hdef}). The remaining level shift in the second line of Eq.~(\ref{eq:C11}) can be calculated by proceeding for example as described in Refs.~\cite{Knight2} using contour integration and standard quantum optical approximations. Doing so one can show that
\begin{eqnarray} \label{eq:LS7}
\Delta_{\rm mirr} &=& {3 r_a \over 2 \eta_a^2} \, \Gamma_{\rm free}  \, {\rm Im} \left[ {{\rm i} \over 2k_{0} x} {\rm e}^{2{\rm i}k_{0} x} (1-\mu) \right. \nonumber \\
&& \left. - {\rm e}^{2{\rm i}k_{0} x} \left( {1 \over (2k_{0} x)^{2}} + {{\rm i} \over (2k_{0} x)^{3}} \right) (1+\mu) \right] ~~~~~~
\end{eqnarray}
which equals $\Delta_{\rm mirr}$ in Eq.~(\ref{eq:rho12}).

\section{Calculation of ${\cal L}(\rho_{\rm I} (t))$ for a semi-transparent mirror} \label{appD}

Substituting Eq.~\eqref{eq:E(x)45} into Eq.~\eqref{eq:rho4xxx}, one can moreover show that ${\cal L} (\rho_{\rm AI} (t))$ equals
\begin{eqnarray} \label{eq:rho100}
{\cal L} (\rho_{\rm AI} (t)) &=&
\frac{1}{\Delta t} \int \limits_{t}^{t + \Delta t} {\rm d}t^{\prime} \int \limits_{t}^{t+\Delta t} {\rm d}t^{\prime \prime} \int_{\mathbb{R}^{3}} {\rm d}^3 {\bf k} \, \sum_{\lambda = 1,2} | g_{{\bf k} \lambda}  |^2 \nonumber \\
&& \times \left[ {1 + r_a^{ 2} \over \eta_a^2} - {2 r_a \over \eta_a^2} \, \cos \left( {\bf k} \cdot ({\bf r} - \widetilde{\bf r}) \right) + {t_b^{ 2} \over \eta_b^2} \right] \nonumber \\
&& \times {\rm e}^{{\rm i} (\omega - \omega_0) (t' - t'')} \, \sigma^- \, \rho_{\rm AI} (t) \, \sigma^+  \, . ~~~~
\end{eqnarray}
This expression can be simplified using the same approximations as in App.~\ref{appC}. Substituting Eqs.~\eqref{eq:rho88}--\eqref{eq:rho888} into this equation and taking into account that
\begin{eqnarray} \label{eq:rho67}
\int \limits_{t' - (t + \Delta t)}^{t'-t} {\rm d}\xi \, {\rm e}^{{\rm i} (\omega - \omega_0) \xi}
&=& 2 \pi \, \delta (\omega - \omega_0)
\end{eqnarray}
to a very good approximation, finally yields 
\begin{eqnarray} \label{eq:rho9b}
{\cal L} (\rho_{\rm AI} (t)) &=& \Gamma_{\rm mirr} \, \sigma^- \, \rho_{\rm AI} (t) \, \sigma^+  
\end{eqnarray}
with $\Gamma_{\rm mirr}$ given in Eq.~\eqref{eq:rho12}.


\end{document}